\documentclass[aps,prb,showpacs,twocolumn,amssymb,floats,epsfig]{revtex4}
\usepackage{epsfig,psfrag,subfigure}
\usepackage{graphicx,psfrag,subfigure}
\usepackage{color}
\usepackage{amssymb,amsbsy,amsmath}
\usepackage{amsmath}

\newcommand\beq{\begin{equation}}
\newcommand\eeq{\end{equation}}
\newcommand\bea{\begin{eqnarray}}
\newcommand\eea{\end{eqnarray}}
\newcommand\al{\alpha}
\newcommand\be{\beta}
\newcommand\ga{\gamma}
\newcommand\de{\delta}
\newcommand\ep{\epsilon}
\newcommand\De{\Delta}
\newcommand\si{\sigma}

\newcommand\la{\lambda}

\newcommand\ta{\theta}
\newcommand\dg{\dagger}
\newcommand\pa{\partial}
\newcommand\non{\nonumber}
\newcommand\noi{\noindent}

\newcommand\ig{\includegraphics}
\newcommand\bib{\bibitem}

\begin{document}

\title{Majorana modes and transport across junctions of superconductors and
normal metals}

\author{Manisha Thakurathi, Oindrila Deb and Diptiman Sen}
\affiliation{Centre for High Energy Physics, Indian Institute of Science, 
Bengaluru 560 012, India}

\date{\today}

\begin{abstract}
We study Majorana modes and transport in one-dimensional systems with 
a $p$-wave superconductor (SC) and normal metal leads. For a 
system with a SC lying between two leads, it is known that 
there is a Majorana mode at the junction between the SC and 
each lead. If the $p$-wave pairing $\De$ changes sign or if a strong impurity 
is present at some point inside the SC, two additional Majorana 
modes appear near that point. We study the effect of all these modes on the 
sub-gap conductance between the leads and the SC. We derive 
an analytical expression as a function of $\De$ and the length $L$ of the 
SC for the energy shifts of the Majorana modes at the junctions 
due to hybridization between them; the shifts oscillate and decay
exponentially as $L$ is increased. The energy shifts exactly match the 
location of the peaks in the conductance. Using bosonization and the 
renormalization group method, we study the effect of interactions between 
the electrons on $\De$ and the strengths of an impurity inside the 
SC or the barriers between the SC and the leads; 
this in turn affects the Majorana modes and the conductance. Finally we 
propose a novel experimental realization of these systems, in particular of a 
system where $\De$ changes sign at one point inside the SC.
\end{abstract}

\maketitle

\section{Introduction}

Topological phases of quantum systems have been extensively studied in 
recent years~\cite{hasan,qi}. Typically, systems in such phases are gapped
in the bulk but have gapless modes at the boundary, and the energy of the 
boundary modes lie in the bulk gap. Further, the number of species of boundary 
modes is given by a topological invariant which is robust against small 
amounts of disorder, and physical quantities (such as conductance) often 
take quantized values. 

A prototypical example of a system with topological phases is the Kitaev 
chain~\cite{kitaev1}; this is a spin-polarized $p$-wave superconductor (SC) 
in one dimension, and it is known to have a zero energy Majorana mode at each 
end of a long system in the topological phase. (It also has a non-topological 
phases in which there are no Majorana modes at the ends). This system and 
others similar to it have been theoretically studied from various points of 
view~\cite{beenakker,lutchyn1,oreg,fidkowski1,fidkowski2,potter,law,fulga,
stanescu1,tewari1,gibertini,lim,tezuka,egger,ganga,stoud,sela,lobos,pientka1,
stanescu2,lutchyn2,roy,cook,pedrocchi,sticlet1,jose,klino1,klino2,sticlet2,
klino3,alicea1,alicea2,stanescu3,chung,shivamoggi,dirks,adagideli,stanescu4,
erik,mano,sau1,sarma,akhmerov,degottardi1,jiang,degottardi2,manisha,niu,
leijnse,pientka2,sau2,liu,lang,brouwer,vish,cai,kesel,ojanen,lucig,lobos2,
kashuba,guigou,door,hassler,ivar,nadj,kao,hegde,dumi,tewari2,spanslatt}, 
and it has inspired several experiments to look for Majorana 
modes~\cite{kouwenhoven,deng,rokhinson,das,finck,lee,finck2,nadj2}.
The experimental signatures are a zero bias conductance peak 
~\cite{kouwenhoven,deng,das,finck,lee,finck2,nadj2}
and the fractional Josephson effect~\cite{rokhinson}. The zero bias peak 
occurs because the Majorana mode (which lies at zero energy for a long enough 
system) facilitates the tunneling of an electron from the normal metal (NM) 
lead into the superconductor.

The Kitaev chain can be generalized in many ways; some of these 
generalizations give rise to more than one Majorana mode at each end of the 
system~\cite{fidkowski1,potter,law,niu,manisha}. The case of additional
Majorana modes appearing in the {\it bulk} (rather than at the end) of the 
system is less well studied. (Strictly speaking, the term Majorana refers
to modes with exactly zero energy. However, for the convenience of notation, 
we will use the term more generally to refer to states which are localized at 
some point in the SC, have an energy which lies in the superconducting gap,
and smoothly turn into Majorana states with exactly zero energy if the wire 
is so long that two such modes cannot hybridize with each other).
It would be interesting to study the effect of such Majorana modes occurring
inside the system on the electronic transport across the system. The effect 
of interactions between the electrons is also of interest. In one dimension, 
short-range density-density interactions are known to have a dramatic effect, 
turning the system into a Tomonaga-Luttinger liquid. The fate of the Majorana 
modes in the presence of interactions is therefore an interesting subject of 
study~\cite{ganga,sela,lutchyn2,lobos,fidkowski2,hassler,stoud,mano,kashuba}.

In this paper, we will study Majorana modes and the charge conductance of 
one-dimensional systems in which a $p$-wave superconductor lies between 
two normal metal leads; we will refer to this as a NSN system. As discussed 
below, we will consider two kinds of SC; in the first case, the $p$-wave 
pairing amplitude $\De$ will be taken to have the same value everywhere in 
the SC, while in the second case, $\De$ will be taken to have a change in sign 
at one point which lies somewhere inside the SC~\cite{spanslatt}. We will see 
that the number of Majorana modes is generally different in the two cases. 
In the first case, there are typically only two Majorana modes (one at each 
end of the SC), while in the second case, two additional Majorana modes appear
near the point there $\De$ changes sign. While this has been pointed out 
earlier~\cite{alicea1,kesel,ojanen,lucig,klino3}, the effect of these 
additional modes on the conductance has not been studied earlier. Such 
additional Majorana modes also appear in the first case if there is a 
sufficiently strong impurity at one point in the SC.

Turning to the conductance, we note that electronic transport across a 
junction of a NM and a SC has been extensively studied for many 
years~\cite{blonder,kasta,tanaka95,hayat,ks01,ks02,kwon,yoko,hu,tanaka2}
and the junction between a topological insulator and a SC has also been
studied~\cite{law2,soori}.
The presence of a SC means that there will be both normal reflection and 
transmission and Andreev reflection and transmission~\cite{andreev,blonder}. 
As a result, we will show that
%Therefore we will find two kinds of current 
%across this NSN system: (i) an electron from first NM lead 
%couples with an electron from the other N and forms a Cooper pair 
%inside S, (ii) an electron or hole from first N tunnels to other N 
%via S by quantum coherent tunneling \cite{blonder}. Along with these two 
%current one can get contribution from local normal and Andreev processes. 
there are two differential conductances which can be measured in this NSN 
system: a conductance from the left lead to the right lead which we will call 
$G_N$, and a Cooper pair conductance from the left lead to the SC which we 
will call $G_C$. For a continuum model of this system, we will first present
the boundary conditions at the junctions between the SC and the NM
which follow from the conservation of both the probability and the charge 
current. Using these boundary conditions, we will analytically
and numerically calculate $G_N$ and $G_C$ as functions of the energy $E$ of 
the electron incident from the left lead and the length $L$ of the SC
for the case where $\De$ has the same value everywhere in the SC.
This conductance calculation will be followed by the discussion of a 
superconducting box with hard walls; we will see that this is an
analytically tractable problem which sheds light on the way the conductance 
varies with $L$. Next, we will discuss the case where $\De$ changes sign at 
one point inside the SC so that it has opposite signs on the two parts of 
the SC. Analytical calculations are difficult in this case; however we will 
numerically calculate $G_N$ and $G_C$ as functions of $E$ and $L$. We will 
then use a lattice model to numerically confirm the appearance
of two additional Majorana modes which lie near the point where $\De$ changes 
sign by calculating the Majorana wave functions and the particle density.
Finally, we will show that even if $\De$ is constant everywhere in the SC,
the presence of an impurity at one point in the SC can give rise to two
Majorana modes near that point. 

Next, we study what happens when there are
interactions between the electrons. We use bosonization and the
renormalization group (RG) method to study a SC with interacting electrons
when there is an impurity at one point in the SC and also impurities 
(barriers) at the ends of the SC. We will study what happens 
to the impurity strengths and the Majorana modes as the length $L$ of the SC 
is varied; from this we will deduce how the conductance varies with $L$. 

The plan of the paper is as follows. In Sec. \ref{sec:model}, we introduce the
continuum model for the NSN system. We then derive the boundary conditions at 
the junctions between the SC and the NM and show how this can be used to 
derive the differential conductances $G_N$ and $G_C$. In Sec. \ref{sec:num}, 
we numerically calculate $G_N$ and $G_C$. In Sec. \ref{sec:ana}, we calculate 
the forms of $G_N$ and $G_C$ for various parameters 
of the system such as the pairing amplitude $\De$ and the length $L$ of the 
SC. We then consider a superconducting box with hard walls to understand how 
the energies of the Majorana modes vary with $L$. In Sec. \ref{sec:delch}, we 
use both a continuum and a lattice model to study the Majorana modes and 
the conductances when $\De$ changes sign at some point in the SC. 
In Sec. \ref{sec:rg}, we use the formalisms of bosonization and RG to show how 
the pairing amplitude $\De$ and the strength of a single impurity inside the 
SC vary with the length $L$; this will be followed by a discussion of the 
effect of the RG variation on the Majorana modes near that point. In Sec. 
\ref{sec:expt}, we will discuss how to experimentally construct the various 
models discussed in the earlier sections. We will end in Sec. \ref{sec:concl} 
with a summary of our main results and some additional comments.

In brief, the main aim of this paper is to consider a simple model of 
a $p$-wave superconducting wire and to study the Majorana modes and 
conductances using analytical techniques as far as possible. 
We have compared the cases of the $p$-wave pairing having
the same sign everywhere in the superconductor and changing sign at
one point in the superconductor (when two additional symmetry
protected Majorana modes appear around that point), to see if there
is an appreciable difference in the conductances in the two cases.
We have proposed experimental set-ups where these two cases can be studied.
We have studied the effect of interactions between the electrons on the 
various parameters of the system and therefore on the conductances. We 
believe that it is useful to have a unified and analytical understanding 
of all these aspects of this important subject. We have not attempted to 
carry out extensive numerical calculations as has been done in many other 
papers.

\section{Model for a NSN system}
\label{sec:model}

\begin{figure}[h]
%\begin{center}\ig[width=2.5in]{NSN1.ps} 
\begin{center}\ig[width=2.5in]{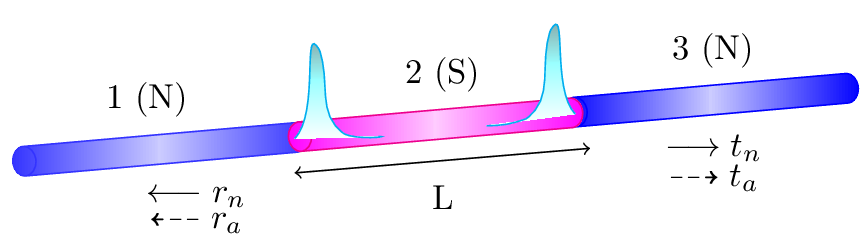} 
\caption{(Color online) Schematic picture of a NSN system. The middle part 
(2) with length $L$ is the $p$-wave superconductor, while the left and right 
parts (1 and 3) are normal metal leads. Four amplitudes are shown: $r_n, ~r_a$
are normal and Andreev reflections in the left lead, and $t_n, ~t_a$ are normal
and Andreev transmissions in the right lead. Majorana modes at the two ends of
the SC region are also shown.} \label{fig:nsn1} \end{center} \end{figure}

We begin with a continuum model for a NSN system in one dimension as shown
in Fig.~\ref{fig:nsn1}. We assume that the NM lead on the left goes from $x
=-\infty$ to 0, while the NM lead on the right goes from $L$ to $\infty$; a 
spin-polarized (hence effectively spinless) $p$-wave superconductor lies in 
the region $0 < x < L$. 

Let us denote the wave function in each region as $\psi= (c,~ d)^T$, where 
$c(x,t), ~d(x,t)$ are the electron (particle) and hole components respectively
(there is no spin label to be considered here). The Hamiltonian in each region 
can be written as
\bea H &=& \int dx~[ c^\dg (- \frac{\hbar^2 \pa_x^2}{2m} -\mu ) c ~-~ d^\dg (- 
\frac{\hbar^2 \pa_x^2}{2m} - \mu ) d \non \\
&& ~~~~~~~~~- \frac{i\De}{k_F} ~( c^\dg \pa_x d + d^\dg \pa_x c ) ], 
\label{Ham} \eea
where $\mu$ is the chemical potential, $k_F = \sqrt{2 m \mu}/\hbar$ is the 
Fermi wave number, and $\De$ is the $p$-wave superconducting pairing amplitude
assumed to be real everywhere; we will set $\De=0$ in the NM leads. (We will
generally set $\hbar = 1$ in this paper, except in places where it is required
for clarity. The Fermi velocity is $v_F = \hbar k_F/m$). The Heisenberg 
equations of motion $i \pa_t c = - [H,c]$ and $i \pa_t d = - [H,d]$ imply that
\bea i \pa_t c &=& (- \frac{\pa_x^2}{2m} -\mu)~c ~-~\frac{i\De}{k_F} ~\pa_x d,
\non \\
i \pa_t d &=& (\frac{\pa_x^2}{2m} +\mu)~d ~-~ \frac{i\De}{k_F} ~\pa_x c. 
\label{eom} \eea
For a wave function which varies in space as $e^{ikx}$, the energy is given 
by $\pm (k^2/(2m) - \mu)$ if $\De = 0$, and by $\pm \sqrt{(k^2/(2m) - \mu)^2 + 
\De^2 (k/k_F)^2}$ if $\De \ne 0$. The corresponding wave functions will be 
presented below. We see that the energy spectrum in the SC has a gap equal to 
$2\De$ at $k= \pm k_F$.

Let us define the particle density $\rho_p=c^\dg c +d^\dg d$ and charge 
density $\rho_c= c^\dg c -d^\dg d$. Using Eqs.~\eqref{eom} and the equations 
of continuity $\pa_t {\rho_p} + \pa_x J_p = 0$ and $\pa_t {\rho_c} + \pa_x J_c
= 0$, we find the particle and charge currents to be~\cite{blonder,soori}
\bea J_p &=& \frac{i}{2m} ~[-~c^\dg\pa_x c ~+~ \pa_x c^\dg c ~+~ d^\dg \pa_x 
d ~-~ \pa_x d^\dg d] \non \\
&& ~+ ~\frac{\De}{k_F} ~(c^\dg d + d^\dg c), \non \\
J_c &=& J_1 ~+~ \int_0^x dx' ~J_2 (x'), \non \\
J_1 &=& \frac{i}{2m} ~[- ~c^\dg \pa_x c ~+~ \pa_x c^\dg c ~-~ d^\dg \pa_x d ~+~
\pa_x d^\dg d] \non \\
&& ~+ ~\frac{\De}{k_F} ~(c^\dg d + d^\dg c)], \non \\
J_2 &=& - ~\frac{2\De}{k_F} ~(\pa_x c^\dg d + d^\dg \pa_x c). \eea
The last term, $J_2$, can be interpreted as the contribution of Cooper pairs 
to the charge current; note that it vanishes in the NM where $\De = 0$.

The boundary conditions at the SC-NM junctions at $x=0$ and $L$ can be found 
by demanding that the currents $J_p$ and $J_c$ be conserved at those points. 
At the junction $x=0$, let us consider the wave functions $\psi_1=(c_1,~d_1)^T$
and $\psi_2= (c_2, ~d_2)^T$ at the points $x=0-\ep$ and $x=0+\ep$, i.e., in 
the NM and SC regions respectively. The condition $J_{p1}(0-\ep)= J_{p2}
(0+\ep)$ implies that 
\bea & & \frac{i}{2m}~[- ~c_1^\dg\pa_x c_1 ~+~ \pa_x c_1^\dg c_1 ~+~ d_1^\dg
\pa_x d_1 ~-~ \pa_x d_1^\dg d_1 ~] \non \\
= & & \frac{i}{2m} ~[-~ c_2^\dg \pa_x c_2 ~+~ \pa_x c_2^\dg c_2 ~+~ 
d_2^\dg \pa_x d_2 ~-~ \pa_x d_2^\dg d_2] \non \\
&& ~+ ~\frac{\De}{k_F} ~(c_2^\dg d_2 + d_2^\dg c_2). \eea
The simplest way of satisfying this condition is to set
\bea c_1&=&c_2, \non \\
d_1&=&d_2, \non \\
\pa_x c_1&=& \pa_x c_2 ~+~ \frac{i\De}{v_F} ~d_2, \non \\ 
\pa_xd_1&=& \pa_x d_2 ~-~ \frac{i\De}{v_F} ~c_2, \label{bc1} \eea
where $v_F = \hbar k_F/m$.
(The first two equations above mean that the wave function is continuous
while the last two equations imply that the first derivative is discontinuous
in a particular way). We now find Eqs.~\eqref{bc1} also imply that that 
charge current is conserved, i.e., $J_{c1}(0-\ep)= J_{c2}(0+\ep)$. 
Next, let us consider what happens if a $\de$-function potential of strength 
$\la$ is also present at the junction at $x=0$; the dimension of $\la$ is
energy times length. (This potential is physically
motivated by the fact that in many experiments, the NM leads are weakly 
coupled, by a tunnel barrier, to the SC. This can be modeled by placing a 
$\de$-function potential with a large strength at the junction).
Now there will be an additional discontinuity in the first derivative
at $x=0$; this is found by integrating over the $\de$-function which gives
\beq \pa_x \psi_2 (0+\ep) ~-~ \pa_x \psi_1 (0-\ep) ~=~ 2m \la ~\psi_1(0). \eeq
Hence Eqs.~\eqref{bc1} must be modified to
\bea c_1 &=& c_2, \non \\
d_1 &=& d_2, \non \\
\pa_x c_1 ~+~ 2m \la ~c_1 &=& \pa_x c_2 ~+~ \frac{i\De}{v_F} ~d_2, \non \\ 
\pa_x d_1 ~+~ 2m \la ~d_1 &=& \pa_x d_2 ~-~ \frac{i\De}{v_F} ~c_2. 
\label{bc2} \eea

For the SC-NM junction at $x=L$, we consider the wave functions $\psi_2= 
(c_2,~ d_2)^T$ at $x=L-\ep$ in the SC region and $\psi_3= (c_3,~ d_3)^T$ 
at $x=L+\ep$ in the NM region. Assuming that there is also a $\de$-function
potential with strength $\la$ at this junction, we find that the boundary
conditions which conserve the probability and charge current at this point 
are given by
\bea c_2 &=& c_3, \non \\
d_2 &=& d_3, \non \\
\pa_x c_2 ~+~ \frac{i\De}{v_F} ~d_2 ~+~ 2m \la ~c_2 &=& \pa_x c_3, \non \\ 
\pa_x d_2 ~-~ \frac{i\De}{v_F} ~c_2 ~+~ 2m \la ~d_2 &=& \pa_x d_3. 
\label{bc3} \eea

We now use the boundary conditions discussed above to find the various
reflection and transmission amplitudes when an electron is incident from,
say, the NM lead on the left with unit amplitude. The presence of the SC
in the middle implies that one of four things can happen~\cite{blonder}.

\noi (i) an electron can be reflected back to the left lead with amplitude
$r_n$.

\noi (ii) a hole can be reflected back to the left lead with amplitude $r_a$. 
In this case, charge conservation implies that a Cooper pair must be produced
inside the SC.

\noi (iii) an electron can be transmitted to the right lead with amplitude 
$t_n$.

\noi (iv) a hole can be transmitted to the right lead with amplitude $t_a$. 
(This is usually called crossed Andreev reflection~\cite{law2}). Then 
charge conservation again implies that a Cooper pair must be produced 
inside the SC.

If the energy $E$ of the electron (incident from the left lead) lies in the 
superconducting gap, i.e., $-\De \le E \le \De$ ($E$ can be interpreted as the 
bias between the left lead and the SC), Eqs.~\eqref{eom} imply that the wave 
functions in the three regions must be of the form
\bea
\hspace*{-0.8cm} \psi_1 &=& e^{ik x}\left( \begin{array}{c}
1 \\
0 \end{array} \right)
+ r_n e^{-ik x} \left(\begin{array}{c}
1 \\
0 \end{array} \right) 
+ r_a e^{ik x} \left(\begin{array}{c}
0 \\
1 \end{array} \right), \non \\
\hspace*{-0.8cm} \psi_2 &=& t_1e^{ik_1x}\left( \begin{array}{c}
1 \\
e^{i\phi} \end{array} \right)
+ t_2 e^{ik_2x} \left(\begin{array}{c}
1 \\
-e^{-i\phi} \end{array} \right) \non \\
& & +~ t_3 e^{ik_3x} \left(\begin{array}{c}
1 \\
e^{-i\phi} \end{array}\right)
+ t_4 e^{ik_4x} \left(\begin{array}{c}
1 \\
-e^{i\phi} \end{array} \right), \non \\
\hspace*{-0.8cm} \psi_3 &=& t_ne^{ik x} \left( \begin{array}{c}
1 \\
0 \end{array} \right)
+ t_a e^{-ik x} \left(\begin{array}{c}
0 \\
1 \end{array} \label{3eq}\right), \label{waves} \eea
where $e^{i\phi}= (E -i\sqrt{\De^2 -E^2})/\De$.
Here $\psi_1$ and $\psi_3$ are the wave functions in the NM leads on the 
left (1) and right (3) respectively, while $\psi_2$ is the wave function in 
the SC region in the middle (see Fig.~\ref{fig:nsn1}). The top and bottom 
entries in the wave functions denote the particle and hole components. The wave
functions in the NM leads are proportional to $e^{\pm ik x}$, where $k$ is 
close to the Fermi wave number $k_F$. In the SC, we have four modes;
two of these decay exponentially and the other two grow as we move from left 
to right. We denote the wave numbers of these by $k_1, ~k_2, ~k_3$ and $k_4$. 
Defining the length scale
\beq \xi ~=~ \frac{\hbar v_F}{\De \sqrt{1-(E/\De)^2}}, \label{xi} \eeq
we find that the decaying modes have $k_1=k_F+i/\xi$ and $k_2=-k_F+i/\xi$, 
while the growing modes have $k_3=k_F-i/\xi$ and $k_4=-k_F-i/\xi$. 

Upon including the $\de$-function potentials with strength $\la$ at the 
junctions at $x=0$ and $x=L$, we get a total of eight boundary conditions 
from Eqs.~\eqref{bc2} and \eqref{bc3} which connect $\psi_1 = (c_1,~d_1)^T$, 
$\psi_2 (c_2,~d_2)^T$ and $\psi_3 = (c_3,~d_3)^T$.
We thus have eight equations for the eight unknowns $r_a, ~r_n, ~t_n, ~t_a, ~
t_1, ~t_2, ~t_3$ and $t_4$. After solving these equations we can calculate
the reflection and transmission probabilities. The conservation law for 
probability current implies that
\beq |r_a|^2 ~+~ |r_n|^2 ~+~ |t_a|^2 ~+~ |t_n|^2 ~=~ 1. \label{prob} \eeq
The net probability of an electron to be transmitted from the left NM lead 
to the right NM lead gives the differential conductance
\beq G_N ~=~ |t_n|^2 ~-~ |t_a|^2. \label{gn} \eeq
The net probability of the electron to be reflected back to the left NM lead is
\beq G_B ~=~ |r_n|^2 ~-~ |r_a|^2. \label{gb} \eeq
The remainder, denoted by the differential conductance $G_C$, is the 
probability of electrons to be transmitted into the SC in the form of 
Cooper pairs. The conservation of charge current implies that
\bea G_C &=& 1 ~-~ G_N ~-~ G_B \non \\
&=& 2 ~(|r_a|^2 ~+~ |t_a|^2), \label{gc} \eea
where we have used Eqs.~\eqref{prob}-\eqref{gb} to derive the last line in
Eq.~\eqref{gc}. The corresponding differential conductances into the right 
lead and the SC are given by $e^2/(2\pi \hbar)$ times $G_N$ and $G_C$, where 
$e$ is the charge of an electron. However, we will ignore the factors of 
$e^2/(2\pi \hbar)$ in this paper and simply refer to $G_N$ and $G_C$ as 
the conductances. 

We note that a differential conductance denotes $G = dI/dV$.
To measure $G_N$ and $G_C$ in our system, we have to assume that there is a 
voltage bias $V$ between the left NM lead on the one hand and the SC and
the right NM lead on the other (the latter two are taken to be at
the same potential). Namely, we choose the mid-gap energy in the SC as zero,
and the Fermi energies in the left and right NM leads as $E=eV$ and zero
respectively. The differential conductances $G_N=dI_N/dV$ and $G_C = dI_C/dV$
are then the derivatives with respect to $V$ of the currents measured in the 
right NM lead and in the SC.

\section{Numerical results for uniform $\De$}
\label{sec:num}

In this section we present our numerical results for $G_N$ and $G_C$ as 
functions of the length $L$ of the SC and the ratio $E/\De$ lying in
the range $[-1,1]$. There is a length scale associated with the SC gap
called %, called as the coherence length 
$\eta = \hbar v_F/\De$. (This is different from the length $\xi$ introduced
in Eq.~\eqref{xi} which depends on the energy $E$). We study three cases, 
namely, $L\ll \eta$, $L\sim \eta$ and $L\gg \eta$.

\begin{widetext}
\begin{center} \begin{figure}[htb]
\subfigure[]{\ig[width=2.2in]{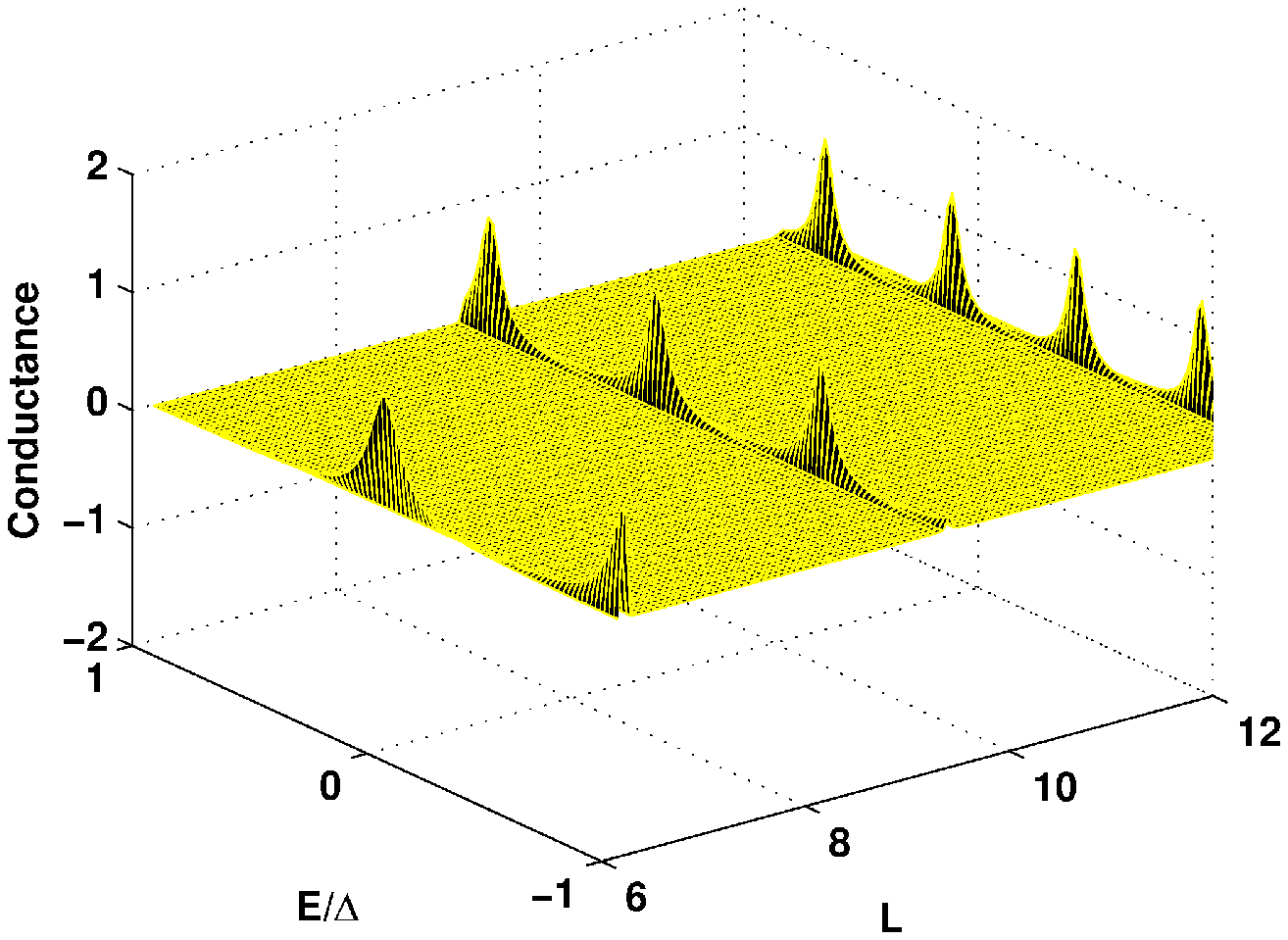}}
\subfigure[]{\ig[width=2.2in]{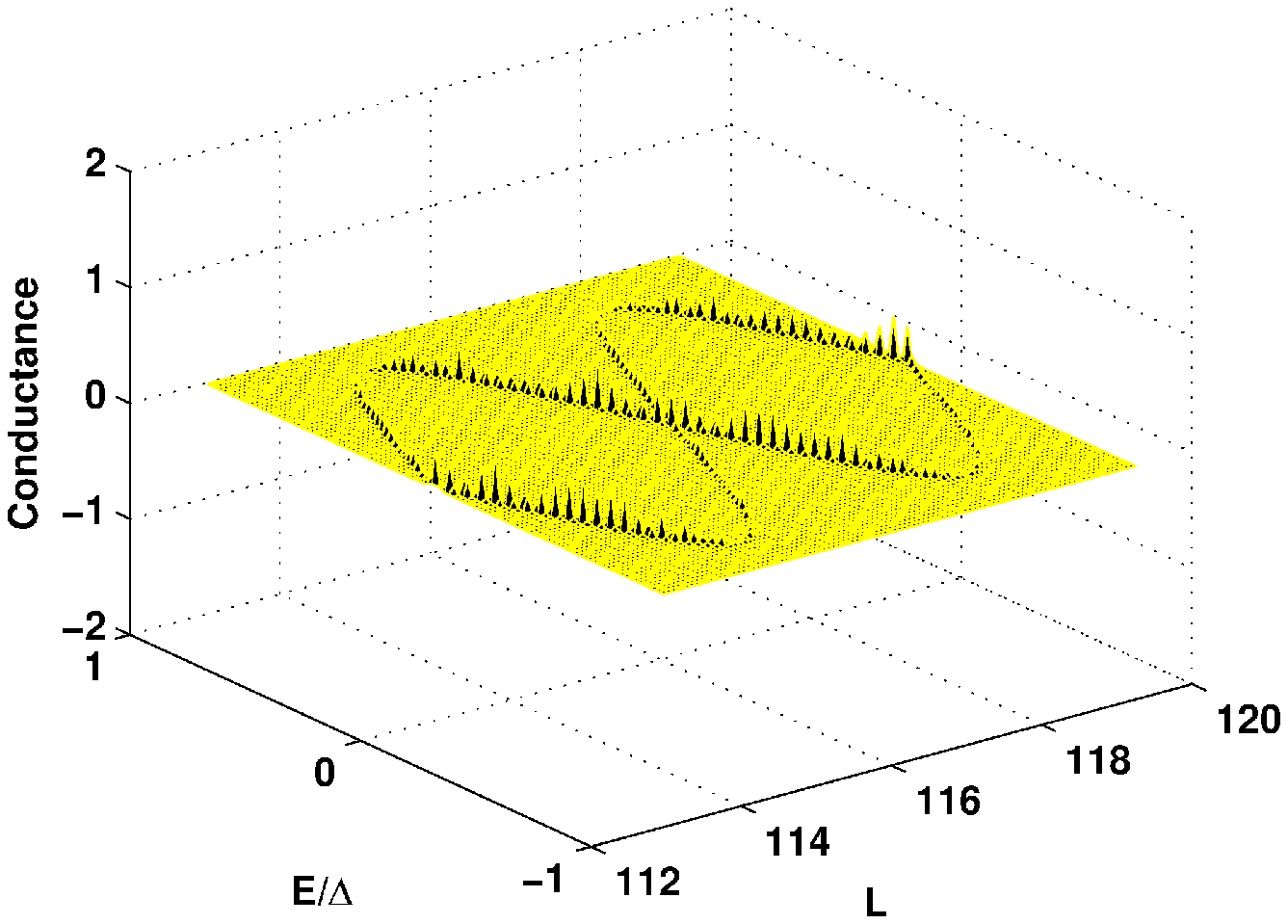}}
\subfigure[]{\ig[width=2.2in]{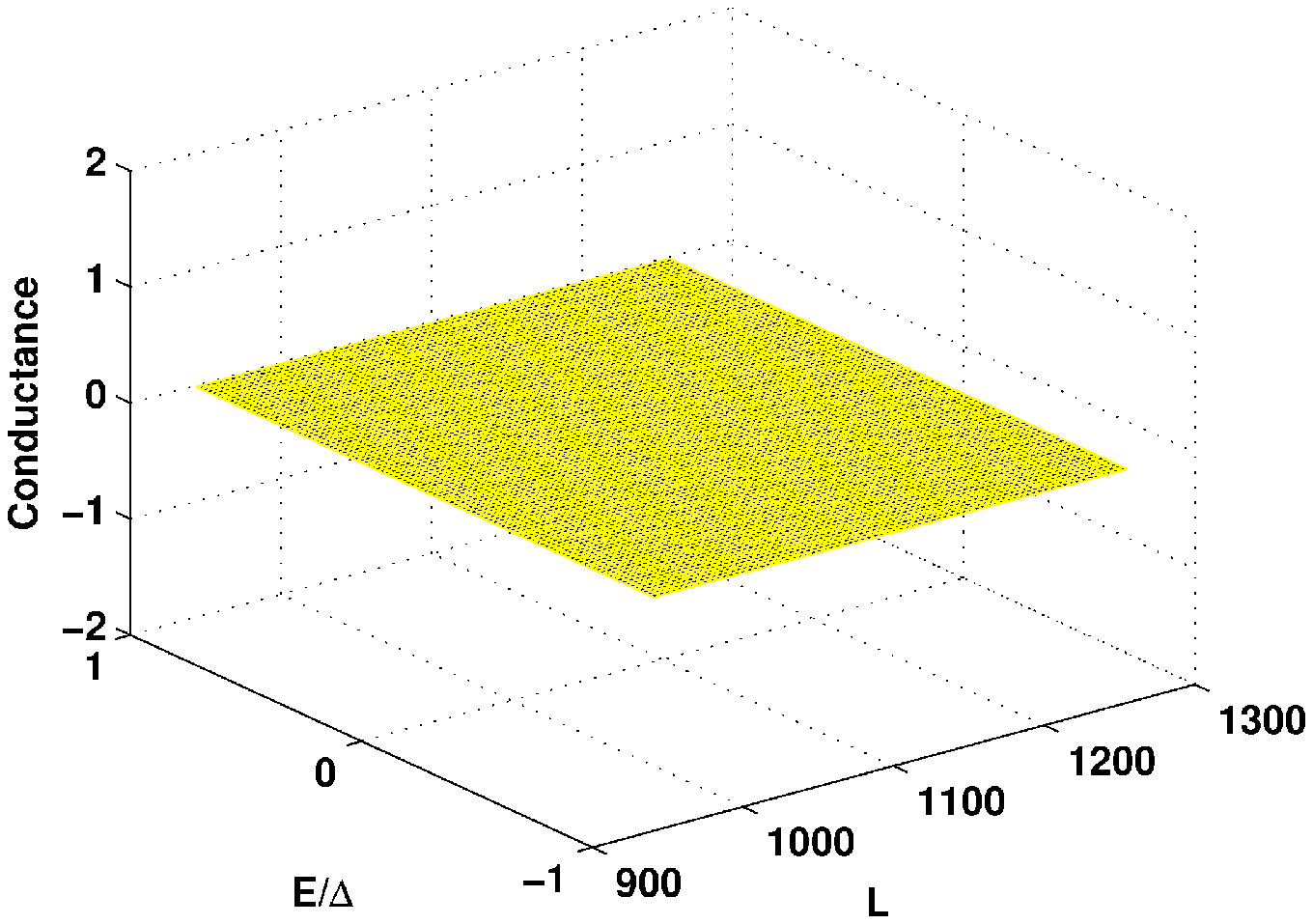}} \\
\subfigure[]{\ig[width=2.2in]{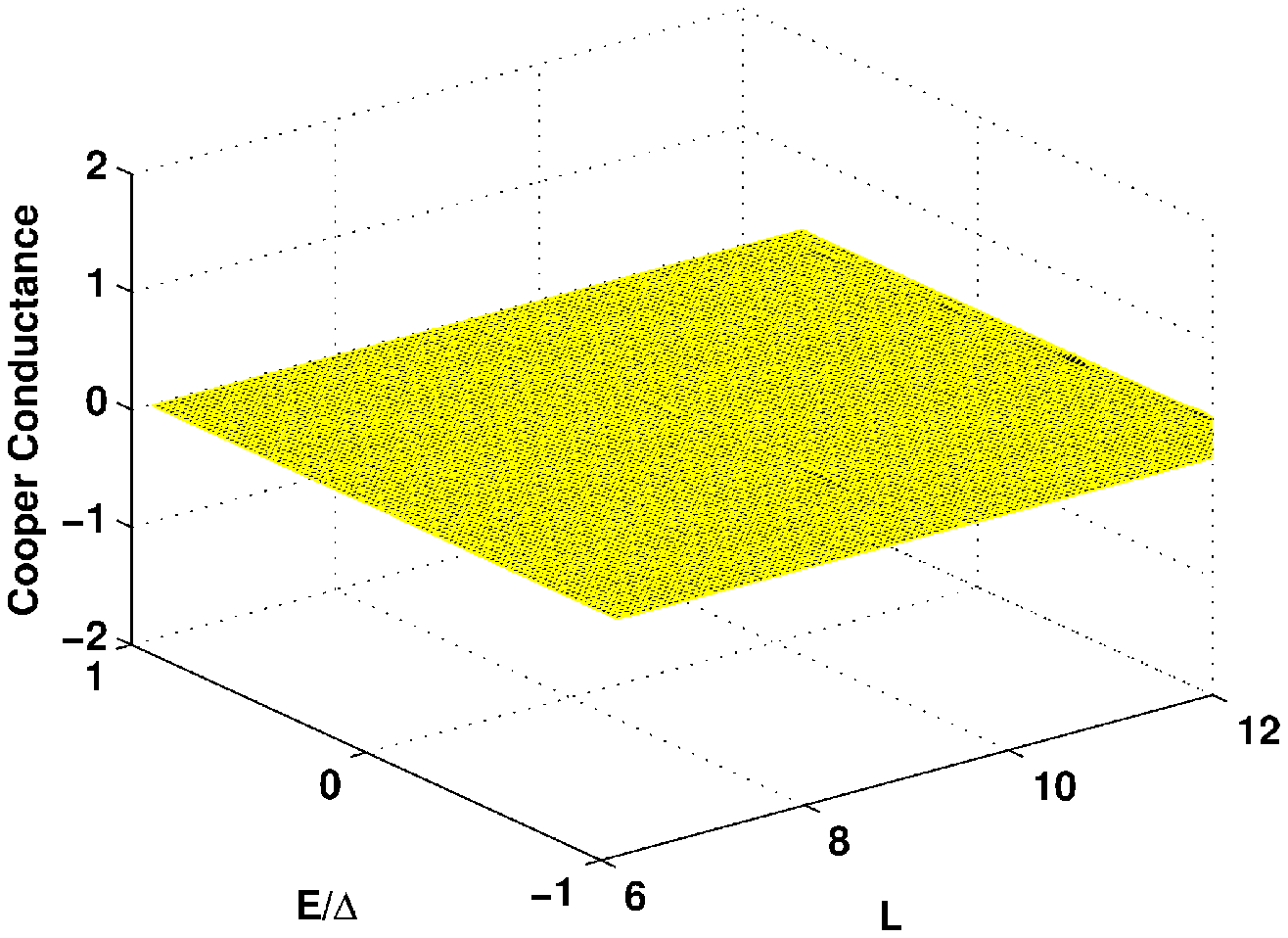}}
\subfigure[]{\ig[width=2.2in]{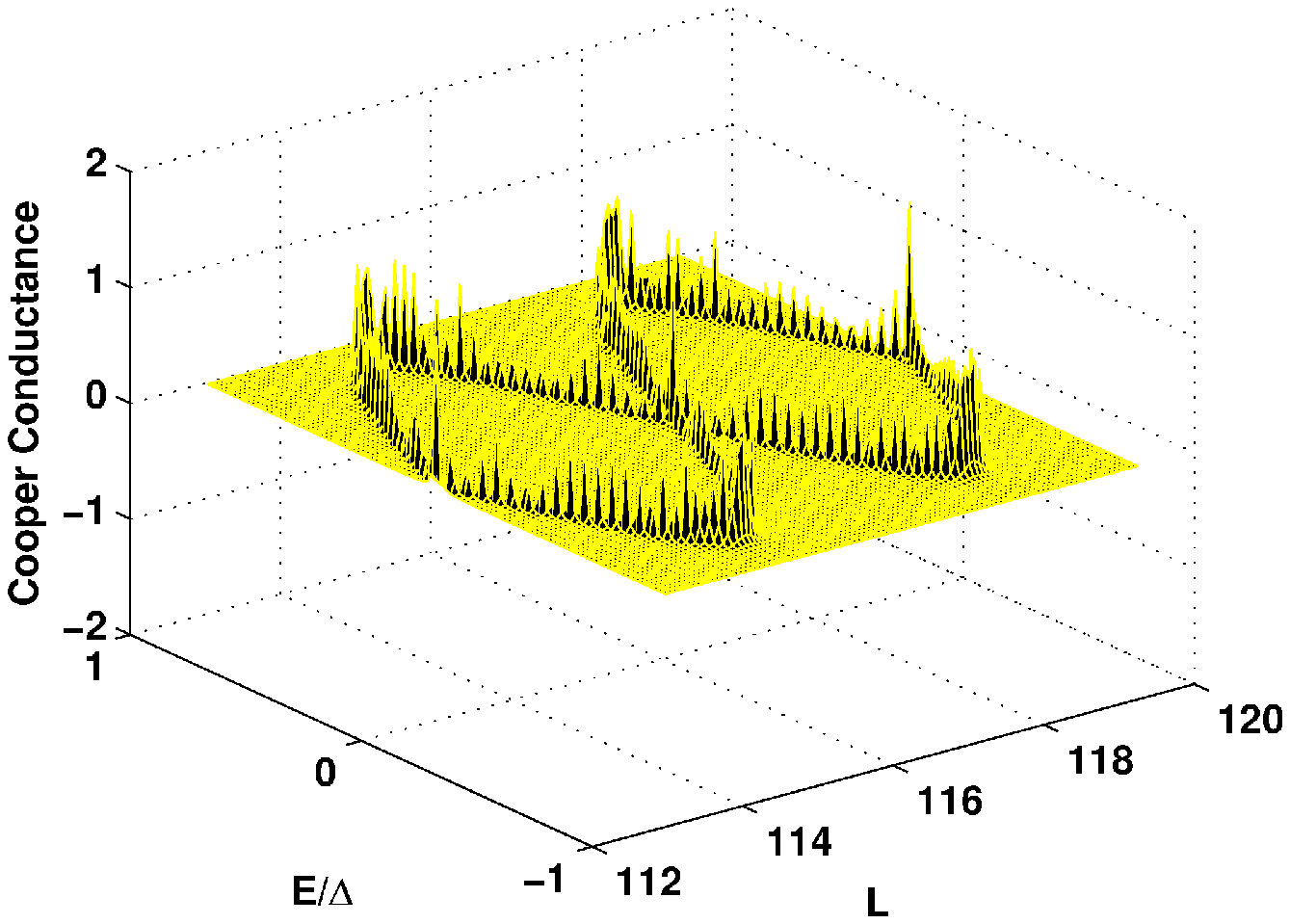}}
\subfigure[]{\ig[width=2.2in]{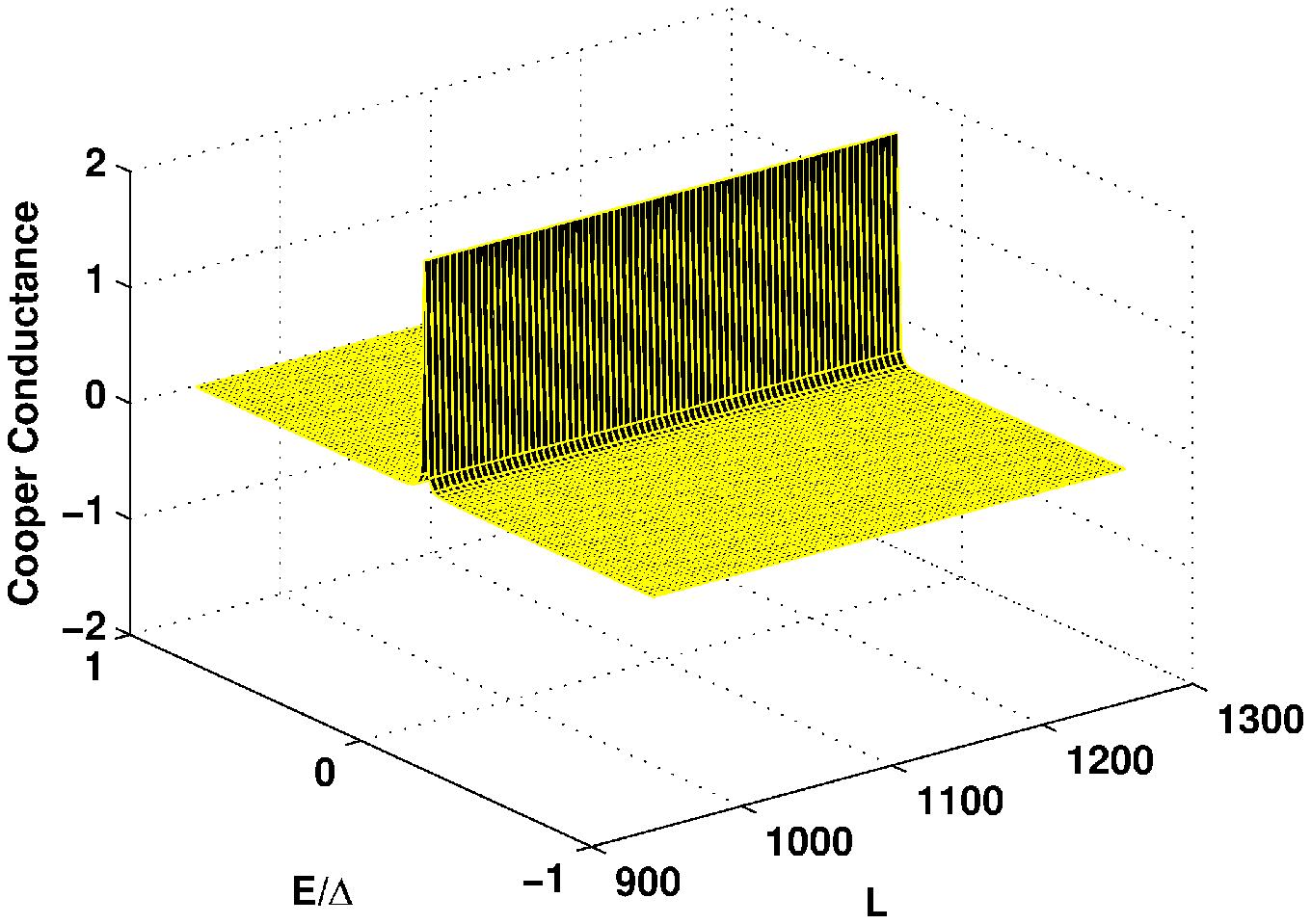}} \\
\subfigure[]{\ig[width=2.2in]{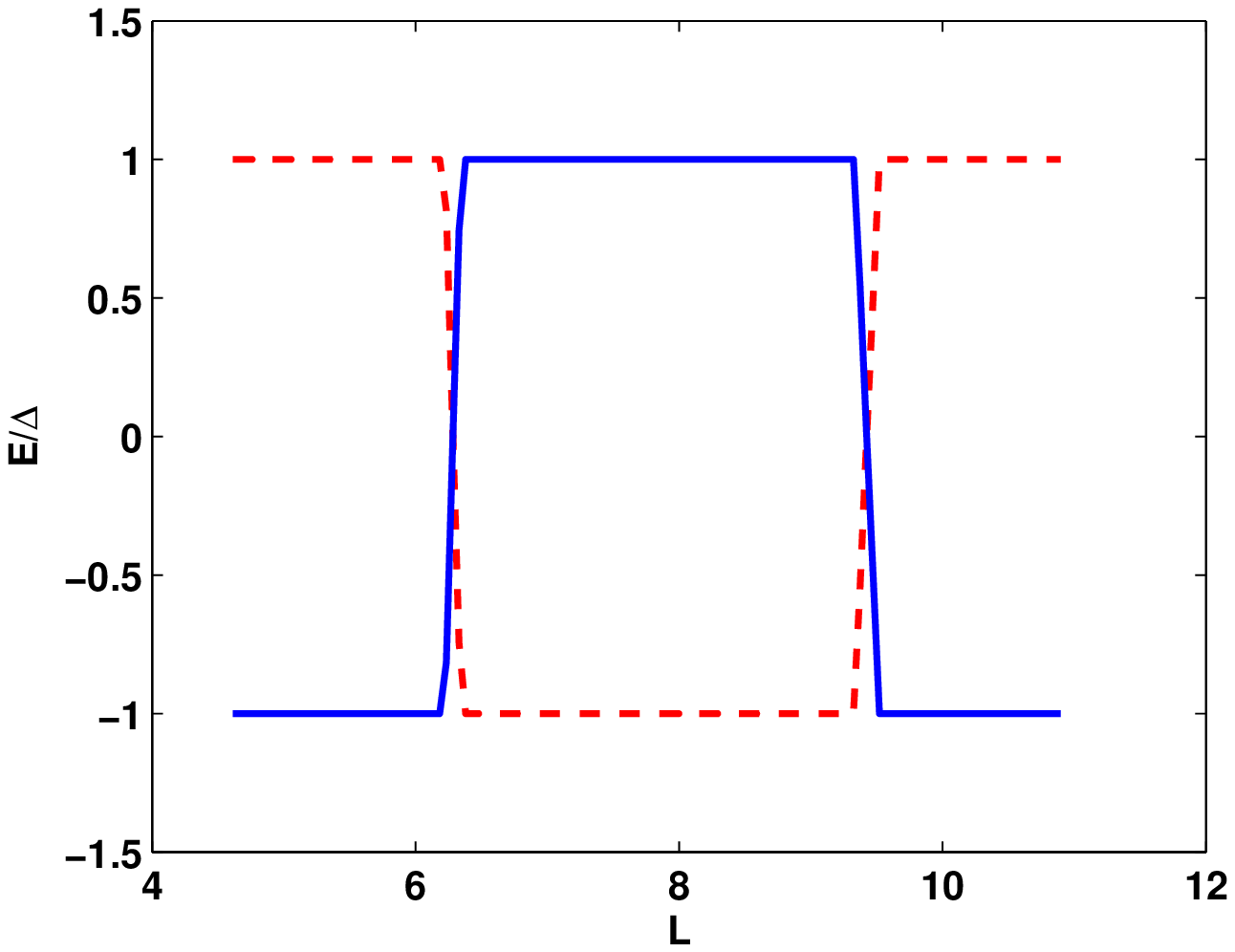}}
\subfigure[]{\ig[width=2.2in]{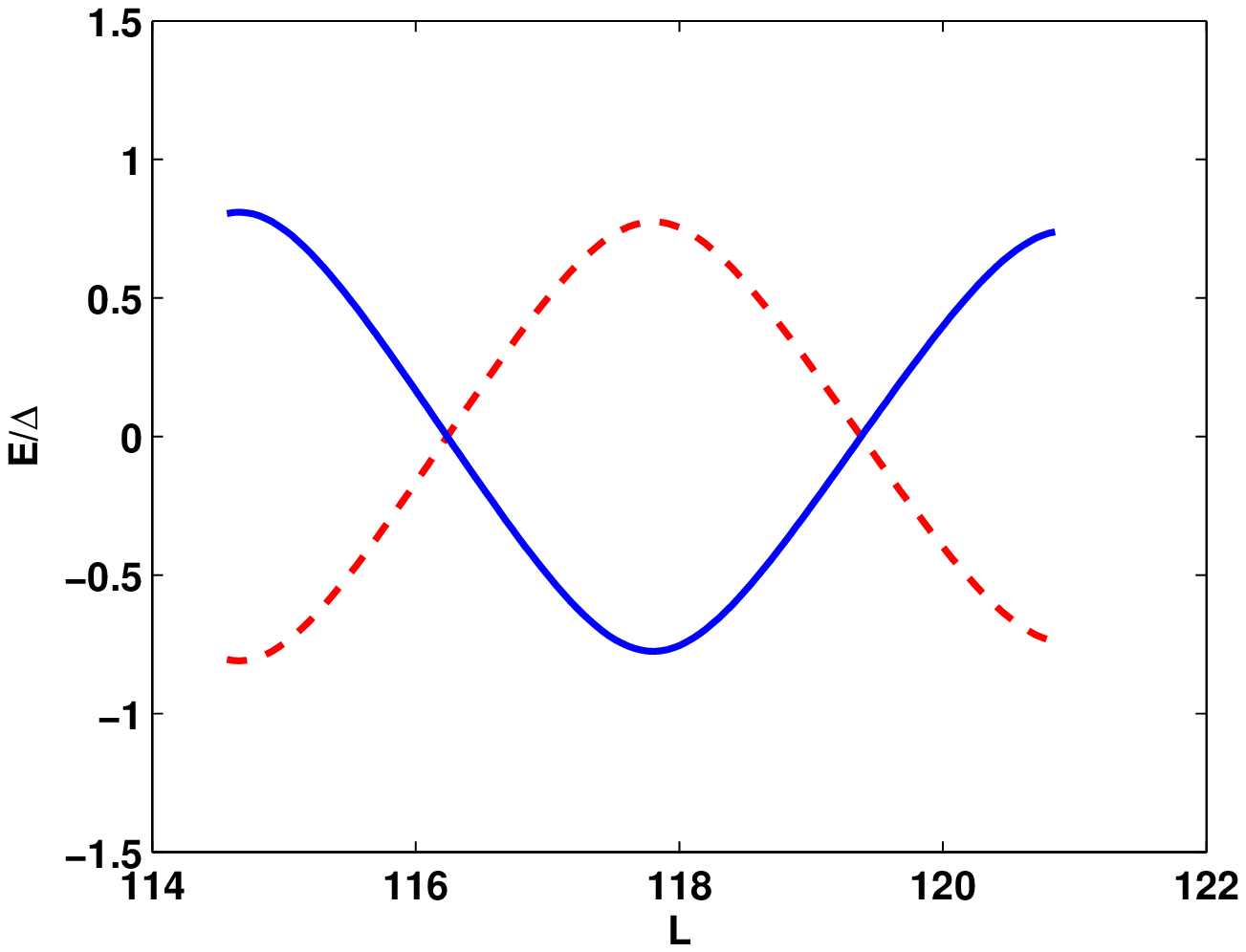}}
\subfigure[]{\ig[width=2.2in]{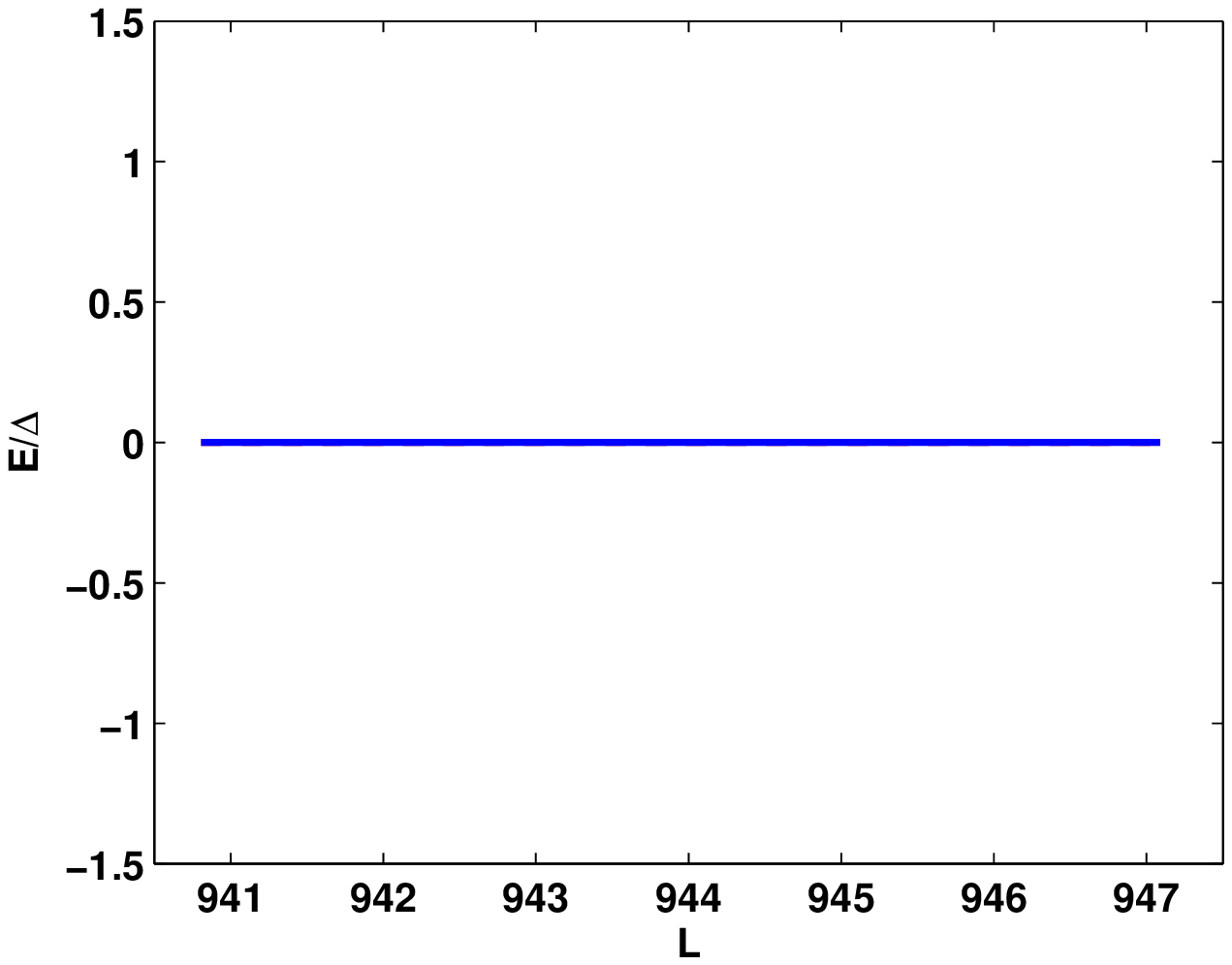}} \\
\caption{(Color online) Plots of conductances and energy splitting when $\De$ 
has the same sign everywhere in the SC: the parameters chosen are $k_F=1, ~m
=1, ~\De=0.01, \la=10$, and an offset in $k_FL$ equal to $0.1$. 
Each row shows three plots corresponding, from left to right, to $L=2\pi-0.1 ~
\text{to}~4\pi-0.1$, $L=37\pi-0.1~\text{to}~ 39\pi-0.1$, and $L=300\pi-0.1~
\text{to}~ 400\pi-0.1$. Figures (a-c) show the conductance $G_N$ from the 
left lead to the right lead while figures (d-f) show the conductance $G_C$ 
from the left lead to the SC, as functions of $L$ and $E/\De$. In figures (a) 
and (d), with $L= 2\pi-0.1 ~\text{to}~ 4 \pi-0.1$, we see that $G_C$ is 
almost zero everywhere, while $G_N$ is a maximum at the values of $L$ where 
the quantization condition $k_F L = n\pi - \text{offset}$ is satisfied. 
(We do not see peaks in $G_N$ at all points on those lines because of the 
finite resolution of the set of values of $L$ and $E/\De$ that we have taken 
in our numerical calculations. As we increase the number of points, we do 
find that the peaks occupy an increasingly large fraction of those lines). 
In figures (b) and (e), with $L= 37\pi -0.1 ~\text{to}~ 39 \pi-0.1$, we see a 
sinusoidal variation in the locations of the peaks of both $G_N$ and $G_C$ 
as a function of $L$. In figures (c) and (f), with $L= 300\pi-0.1 ~\text{to}~ 
400 \pi-0.1$, $G_N$ is almost zero everywhere, while $G_C$ has peaks only 
at $E/\De = 0$ (zero bias peak) where its value is 2. 
Figures (g-i) show the analytically calculated energies $E/\De$ of the 
Majorana modes as a function of the length $L$ of a SC box. The conductance
peaks in figures (a-f) occur exactly at the Majorana mode energies in
the figures (g-i).} \label{fig:con1} \end{figure} \end{center}
\end{widetext}

We show the numerically obtained conductances $G_N$ and $G_C$ in the first 
and second rows of Fig.~\ref{fig:con1}. We have chosen the parameters $k_F=1,~
m=1, ~\De=0.01, ~\la=10$, and an offset in $k_F L$ equal to $0.1$ (the reason 
for the offset is explained in Sec. \ref{sec:box}). The length scale $\eta = 
\hbar^2 k_F/(m\De) = 100$. In the first column of Fig.~\ref{fig:con1}, we have
taken $L= 2\pi-0.1 ~\text{to}~ 4 \pi-0.1$, namely, $L\ll \eta$. We then find 
that $G_N$ is peaked at the values of $L$ where the quantization condition, 
$k_F L = n\pi- \text{offset}$, is satisfied, and $G_C$ is almost zero 
everywhere. For a fixed value of $k_F$, we find that the offset depends only 
on the strength of the $\de$-function potential $\la$ as shown in Eq.
(\ref{offset}) below. In the second column, we have taken $L= 37\pi-0.1 ~
\text{to}~ 39 \pi-0.1$, i.e., $L\sim \eta$. Here we find that 
the locations of the peaks of both $G_N$ and $G_C$ vary sinusoidally with
$L$. In the third column, we have taken $L= 300\pi-0.1 ~\text{to}~ 400 
\pi-0.1$, i.e., $L\gg \eta$. Here $G_C$ has peaks only at zero energy
(called the zero bias peak) where its value is 2, while $G_N$ is almost zero 
everywhere. The general pattern of variation of the conductances is that 
$G_N$ decreases while $G_C$ increases with increasing $L$. For very large $L$,
we observe only a zero bias peak in $G_C$, and only the junction at $x=0$ is 
important; the electron never reaches the junction at $x=L$.

\section{Analytical results for uniform $\De$}
\label{sec:ana}

\subsection{Superconducting box}
\label{sec:box}

In this section, we study how the energies of the Majorana modes depend on 
the length $L$ of the SC region if the NM leads are absent. We consider a SC 
box with hard walls ($\la = \infty$) at $x=0$ and $L$. We therefore set the 
wave functions $\psi_1$ and $\psi_3$ equal to zero in Eq.~\eqref{3eq} and 
consider only the wave function $\psi_2$. This satisfies a total of four 
boundary conditions, two at each wall.

At $x=0$, we obtain
\bea t_1+t_2+t_3+t_4 &=&0, \non \\
t_1 e^{i\phi}-t_2 e^{-i\phi} +t_3 e^{-i\phi}-t_4 e^{i\phi} &=&0 , \eea
while at $x=L$,
\bea t_1 e^{i k_1 L}+t_2 e^{i k_2 L}+t_3 e^{i k_3 L }+t_4 e^{i k_4 L} &=&
0 ,\non \\
t_1 e^{i\phi} e^{i k_1 L }-t_2 e^{-i\phi} e^{i k_2 L } +t_3 e^{-i\phi} 
e^{i k_3 L}-t_4 e^{i\phi} e^{i k_4 L} &=& 0. \non \\ \eea 
The consistency of these four equations implies the condition
\bea \frac{E}{\De} ~=~ \pm ~\frac{\sin(k_F L)}{\sqrt{\sin^2 (k_F L) ~+~ 
\sinh^2 (L/\xi)}}. \label{spl} \eea
This splitting of the energy away from zero is essentially due to the
hybridization of the Majorana modes due to a finite tunneling amplitude
between the two ends of the SC box. We note that oscillations in the 
energy splitting due to the factor of $\sin(k_F L)$ have been studied in 
certain regimes of the wire length in Refs. \onlinecite{sarma}, 
\onlinecite{pientka2}, \onlinecite{klino2} and \onlinecite{stanescu4}.
However, the analytical expression in Eq.~\eqref{spl} is valid for all
values of $L$.

Note that in the limit $L/\xi \to \infty$, the energy splitting goes to zero, 
i.e., $E \to 0$. In this limit, the Majorana modes at the two ends of the box 
are decoupled from each other. The mode at the left end of the system ($x=0$)
has a wave function of the form $(1, ~-i)^T \sin (k_Fx) e^{-x/\xi}$, while 
the mode at the other end $(x=L)$ has a wave function of the form $(1,~i)^T 
\sin (k_F (L-x)) e^{- (L-x)/\xi}$. We have assumed so far that $\De > 0$.
If $\De < 0$, we find that the mode at the left end is proportional to
$(1,~i)^T$ while the mode at the right end is proportional to $(1, ~-i)^T$.

[Incidentally, the wave functions of the Majorana modes can be made completely
real by a unitary transformation. In Eq.~\eqref{Ham}, let us change the phases 
of $c$ and $d$ by $e^{i\pi/4}$ and $e^{-i \pi/4}$ respectively; this 
effectively changes the phase of $\De$ by $i$. We then find that
the Majorana wave functions become $(1,~\pm 1)^T$ instead of $(1,~\pm i)^T$.
Since their energy is zero, there is no complex phase factor $e^{-iEt}$.
So the wave functions are real at all points in space and time].

To visualize the expression in Eq.~\eqref{spl}, we show the dependence of 
$E/\De$ on $L$ in the third row of Fig.~\ref{fig:con1}. For small values of 
$L/\eta \ll 1$, the energies are split by the maximum amount ($E/\De = \pm 
1$), unless $k_F L$ is exactly equal to $n\pi$ where $E/\De = 0$.
As $L$ increases the splitting decreases as we see in Eq.~\eqref{spl} and in
the second plot in the third row of Fig.~\ref{fig:con1}. For $L/\eta \gg 1$,
the energy splitting goes to zero as the tunneling between the two ends
approaches zero exponentially with the length. We see that the analytical 
result in Eq.~\eqref{spl} matches with our numerical results.

We end this section by summarizing our understanding of the conductance 
peaks shown in Fig.~\ref{fig:con1}. In the first column of that figure, the 
length $L$ of the superconducting part of the wire is small. Hence we can 
ignore the imaginary part, $\pm i L/\xi$, in the phase factors appearing in 
$\psi_2$ in Eq.~\eqref{waves}. Thus the wave functions in the SC are 
approximately given by simple plane waves $e^{\pm i k_F x}$. Such a system 
has quantized energy levels when $k_F L = n \pi$; we then get transmission 
resonances, i.e., $|t_n|^2 = 1$ and $G_N =1$. $G_C$ is almost zero since the 
superconducting part is small. In the third column of Fig.~\ref{fig:con1}, 
$L$ is much larger than the Majorana decay length $\xi$. The system therefore
has zero energy Majorana modes near each end of the superconductor; the
two modes are decoupled from each other since the Majorana wave functions
decay exponentially as we go away from the ends, and $L$ is much larger than
$\xi$. In this situation we
find peaks in $G_C$ at zero bias since an electron coming in from the left
lead can enter the superconductor by coupling to the Majorana mode; it then
turns into a Cooper pair and a hole goes back into the left lead to conserve
charge. We thus get perfect Andreev reflection, so that $|r_a|^2 = 1$ and 
$G_C = 2$. $G_N$ is almost zero since the right lead is far away and the
exponential decay of the wave function means the electron cannot reach there.
In the second column of Fig.~\ref{fig:con1}, $L$ has intermediate values of
the same order as $\xi$; the Majorana modes at the two ends can now hybridize 
with each other. We have shown above that the Majorana wave functions 
oscillate as $\sin (k_F x)$ and $\sin (k_F (L-x))$. The hybridization between 
the Majorana modes at the two ends is proportional to the overlap of their 
wave functions and therefore oscillates with the length as $\sin (k_F L)$. 
Hence the energy splitting and therefore the peaks in $G_N$ and $G_C$ also 
oscillate as $\sin (k_F L)$.

\subsection{Conductances}

In this section, we will present the exact forms of $G_N$ and $G_C$ in some 
special cases. It is generally difficult to analytically solve the eight 
equations coming from Eqs.~\eqref{bc2}-\eqref{bc3}. But if $E=0$ 
which implies that $e^{i\phi} =-i$ and $\xi = \hbar v_F/\De$, and if $\De \ll 
\hbar^2 k_F^2/m$, it turns out that one can analytically solve the eight 
equations in two limits.

\noi (i) If $\la/(\hbar v_F) \gg 1$ but $e^{L/\xi}$ has a finite value, the 
numerical calculations of $r_n, ~r_a, ~t_n$ and $t_a$ show that for most
values of the length $L$, $|r_n| \simeq 1$ and the other three amplitudes
almost vanish; hence $G_N, ~G_C$ are very small. However, we find numerically
that there are special values of $L$ where $r_n, ~t_a$ are almost equal to 
zero. We can then use this numerical observation and the boundary conditions 
to analytically find the quantization condition on $L$. We find that
\bea e^{i2k_FL} &=& -~ \frac{v_F + i\la}{v_F - i\la}, \non \\
{\rm implying ~that} ~~k_F L &=& n\pi- \ta, \non \\
{\rm where} ~~\ta &\simeq& \frac{\hbar v_F}{\la}, \non \\
{\rm and} ~~|r_a| &=& \tanh ~(\frac{L}{\xi}), \non \\
|t_n| &=& {\rm sech} ~(\frac{L}{\xi}). \label{offset} \eea
Hence $G_N = {\rm sech}^2 (L/\xi)$ and $G_C = 2 \tanh^2 (L/\xi)$.
We see that if $L/\xi \to 0$, $G_N \to 1$ and $G_C \to 0$, while if 
$L/\xi \to \infty$, $G_N \to 0$ and $G_C \to 2$. (Results similar to
Eqs.~\eqref{offset} have been derived in Ref. \onlinecite{lobos2}). 

Eq.~\eqref{offset} shows that if $\la$ is large but not infinite, there is 
a small offset equal to $\ta$ from the quantization condition $k_F L = n 
\pi$ which is true for $\la = \infty$ (as we see in Eq.~\eqref{spl} for
$E=0$); hence $\sin (k_F L)$ is non-zero but small. For $k_F =1, ~m=1$ and 
$\la= 10$, we see that $\ta = 0.1$, which matches with the numerical results 
shown in Sec. \ref{sec:num}. 

\noi (ii) If $\la/(\hbar v_F)$ and $e^{L/\xi}$ are both much larger than 1 and
$\sin (k_F L)$ is not small (i.e., $k_F L$ is not close to an integer multiple
of $\pi$), we analytically find that
\bea |r_n| &=& \frac{\nu^2}{1 ~+~ \nu^2}, \non \\
|r_a| &=& \frac{1}{1 ~+~ \nu^2}, \non \\
|t_n| &=& |t_a| ~=~ \frac{|\nu|}{1 ~+~ \nu^2}, \non \\
{\rm where} ~~\nu &=& \left( \frac{2\la}{\hbar v_F} \right)^2 ~e^{-L/\xi}~ 
\sin (k_F L). \non \\
{\rm Hence} ~~G_N &=& 0 ~~{\rm and}~~ G_C ~=~ \frac{2}{1 ~+~ \nu^2}. 
\label{nu} \eea
We see that if $\nu \to \infty$, $G_C \to 0$, while if $\nu \to 0$, $G_C \to 
2$. There is a cross-over from one limit to the other depending on whether 
$(\la /\hbar v_F)^2$ (the square of the strength of the barrier between the 
SC and the NM) is much larger than or much smaller than $e^{L/\xi}$.

\section{Effects of $\De$ changing sign}
\label{sec:delch}

\subsection{Continuum model}

\begin{figure}[h]
%\begin{center} \ig[width=2.5in]{NSN1b.ps} 
\begin{center} \ig[width=2.5in]{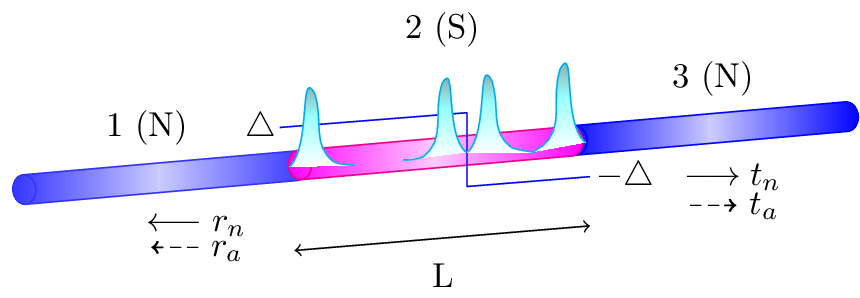} 
\caption{(Color online) Schematic picture of a NSN system where $\De$ 
changes sign at one point in the SC. The middle part with length $L$ is the 
$p$-wave superconductor, while the left and right parts are normal metal 
leads. $r_n, ~r_a$ are normal and Andreev reflection amplitudes in the 
left lead, and $t_n, ~t_a$ are normal and Andreev transmission amplitudes in 
the right lead. Also shown are four Majorana modes, two at the ends of the 
SC region and two near the point where $\De$ changes sign.} \label{fig:nsn1b} 
\end{center} \end{figure}

In this section, we will consider a continuum model for a system in which the 
$p$-wave pairing amplitude $\De$ changes sign at some point in the SC as shown
in Fig.~\ref{fig:nsn1b}. (A way of experimentally realizing such a system will
be discussed in Sec. \ref{sec:expt}). We now get twelve boundary conditions:
four at the NM-SC junction at $x=0$, four at the SC-NM junction at $x=L$, and 
four at the point in the SC where $\De$ changes sign. 

Using the boundary conditions we numerically calculate the conductances. These 
are shown in Fig.~\ref{fig:con2} for the parameter values $k_F=1, ~m=1$, $\la
=10$, and an offset in $k_FL= 0.1$ where $L$ is the length of the SC. We take
$\De=0.01$ from $x=0$ to $3L/4$, and $\De=-0.01$ from $x=3L/4$ to $L$. 
(We found numerically that if $\De$ changes sign exactly at the mid-point of 
the SC, the system has an extra symmetry which gives rise to a rather unusual
pattern of conductances. Since experimentally the sign change will typically
not occur at the mid-point, we chose it to be at $3L/4$ where there is
no special symmetry). As in Sec. \ref{sec:num} where $\De$ was assumed to have 
the same sign everywhere in the SC, we consider three cases. 
For $L= 2\pi-0.1 ~\text{to}~ 3 \pi-0.1$, i.e., $L \ll \eta$, we get 
peaks in $G_N$ at those values of $L$ where the quantization condition 
$k_F L = (n\pi- \text{offset}$) is satisfied; $G_C$ is almost zero. 
For $L= 36\pi-0.1 ~\text{to}~ 38 \pi-0.1$, i.e., $L\sim \eta$, 
we find a sinusoidal variation of the locations of the peaks in both $G_N$ 
and $G_C$ as functions of $L$. For $L= 300\pi-0.1 ~\text{to}~ 400 \pi-0.1$,
i.e., $L\gg \eta$, we find that $G_N$ is almost zero but $G_C$ has peaks
only at zero energy where its value is 2. Thus the conductances show
very similar behaviors as functions of $L$ and $E/\De$ for $\De$ having the 
same sign throughout the SC or changing sign at one point in the SC.

Although the conductances show similar behaviors for systems in which $\De$
has the same sign everywhere or changes sign at one point, we will show below 
that there is a difference in the Majorana mode structure in the two cases. 
Namely, if $\De$ changes sign at one point in the SC, two Majorana modes will 
generally appear near that point. But if $\De$ has the same sign everywhere,
Majorana modes generally do not appear inside the SC unless a large impurity 
potential is present at one point (which has the effect of dividing the
SC into two regions).

\begin{widetext}
\begin{center} \begin{figure}[h]
\subfigure[]{\ig[width=2.2in]{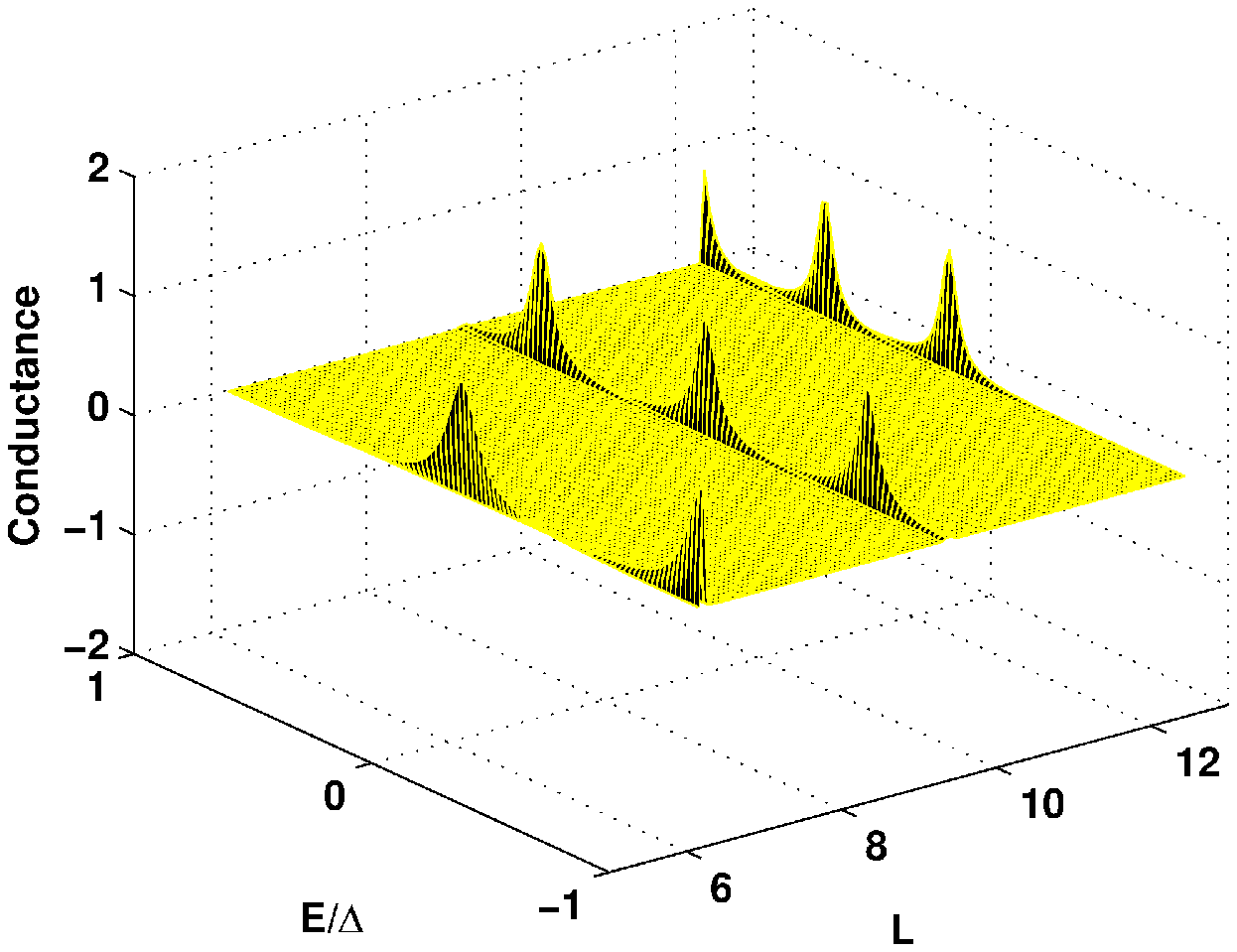}}
\subfigure[]{\ig[width=2.2in]{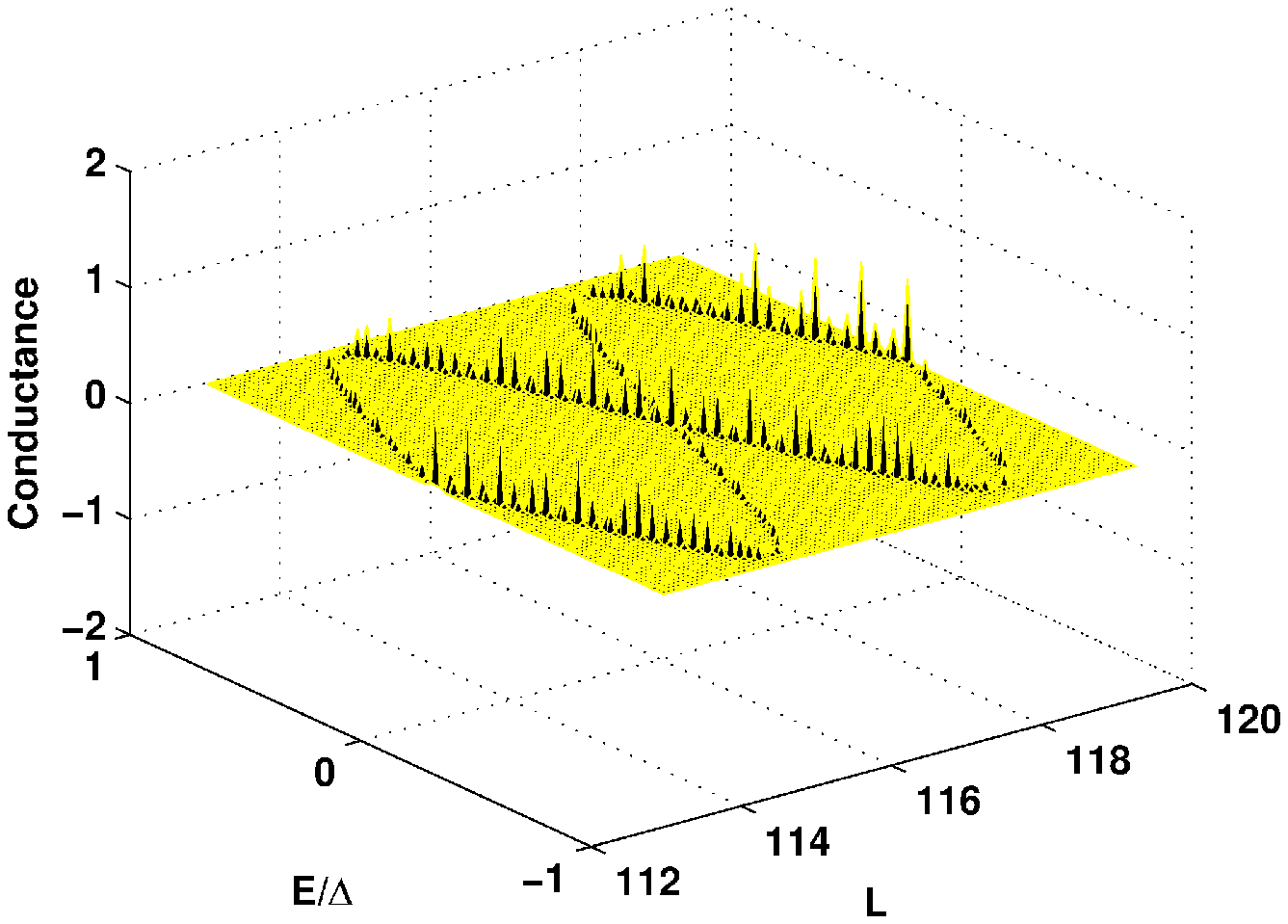}}
\subfigure[]{\ig[width=2.2in]{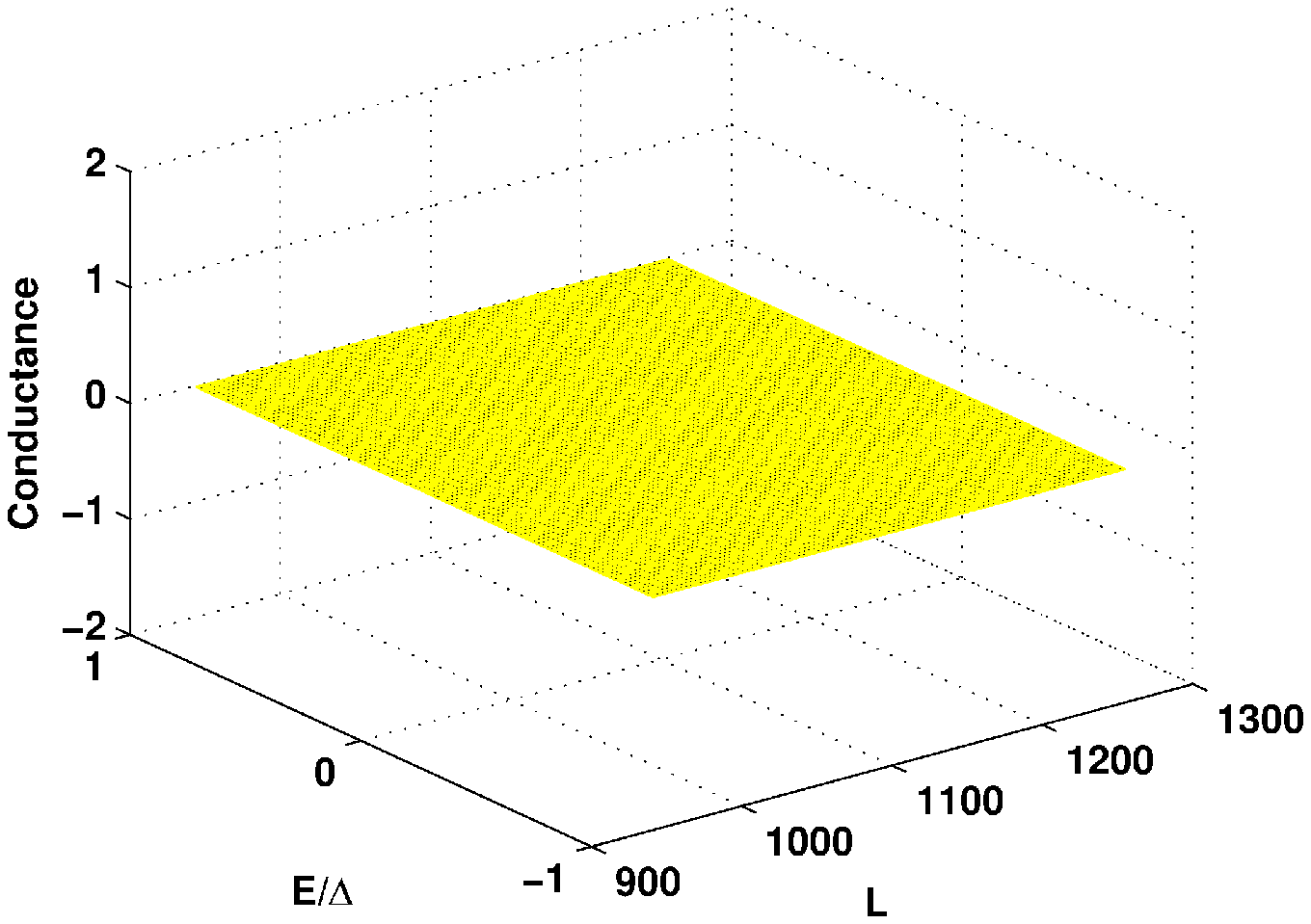}} \\
\subfigure[]{\ig[width=2.2in]{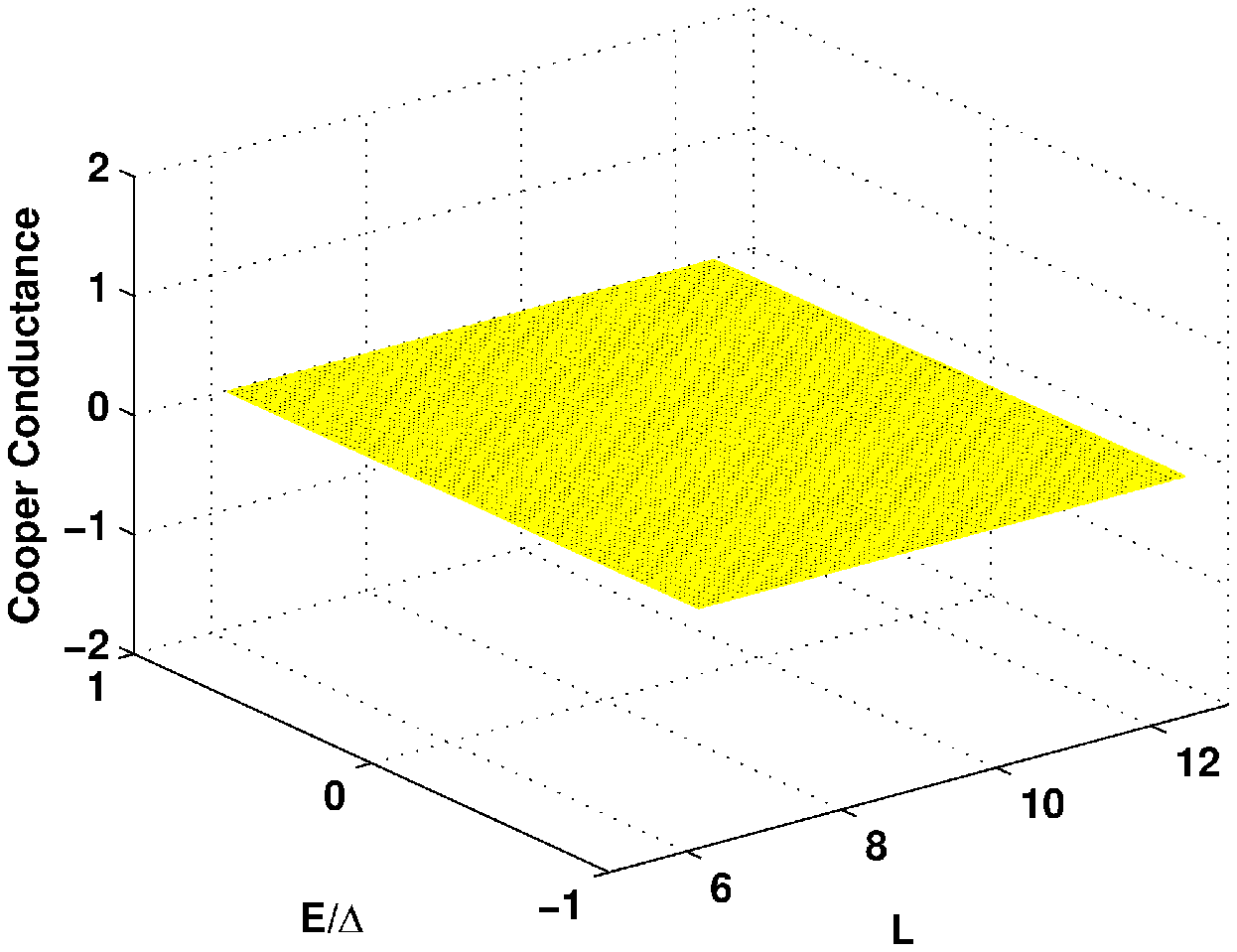}}
\subfigure[]{\ig[width=2.2in]{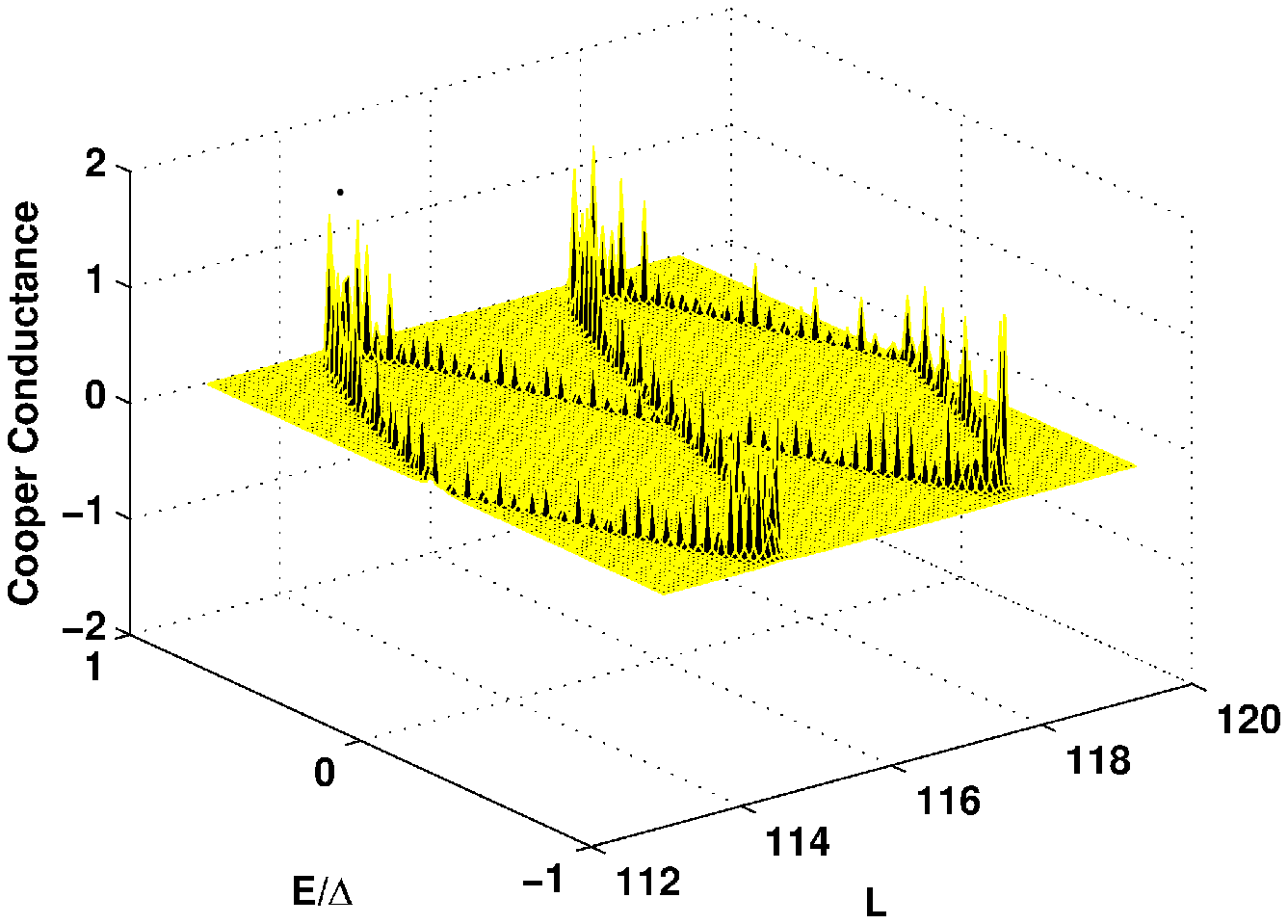}}
\subfigure[]{\ig[width=2.2in]{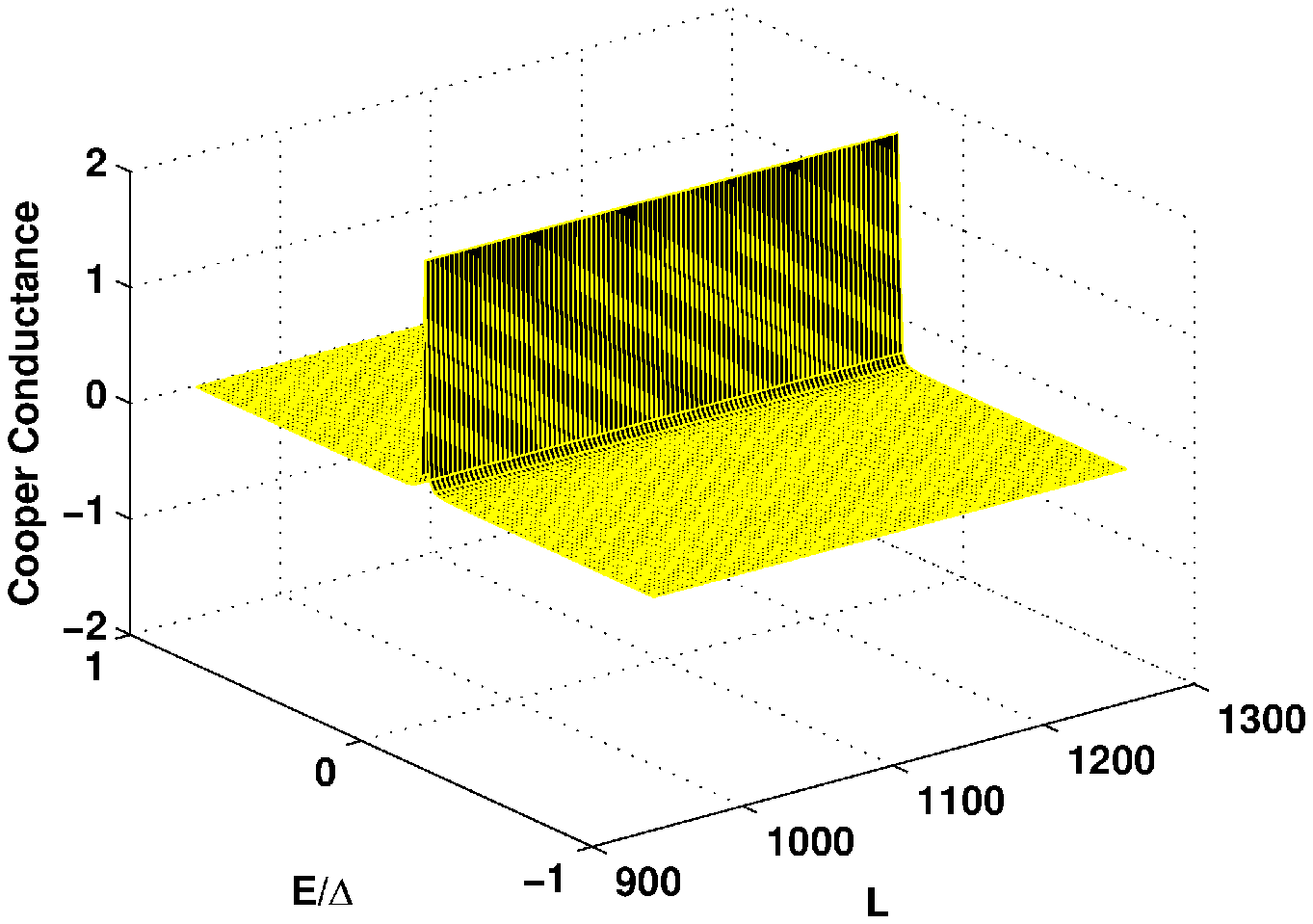}} \\
\caption{(Color online) Plots of conductances when $\De$ changes sign at one 
point in the SC: the parameters chosen are $k_F =1, ~m=1, ~\De = 0.01, ~\la 
= 10$, and an offset in $k_F L$ equal to $0.1$.
The first and second rows respectively show the conductance $G_N$ 
from the left lead to the right lead and the conductance $G_C$ from the
left lead to the SC, as functions of $L$ and $E/\De$. The three different 
plots in the first two rows are for different ranges of $L$. The first column 
shows $L=2\pi-0.1 ~\text{to}~ 4\pi-0.1$, the second column shows 
$L=36\pi-0.1~\text{to}~ 38\pi-0.1$, and the third column shows 
$L=300\pi-0.1~\text{to}~ 400\pi-0.1$.} \label{fig:con2} \end{figure}
\end{center}
\end{widetext}

\subsection{Kitaev chain}

In this section we will obtain a clearer picture of the Majorana modes
by studying the Kitaev chain. This is a lattice model of spinless $p$-wave 
SC with nearest neighbor hopping $\ga$ (which we will assume to be positive), 
SC pairing $\De$, and chemical potential $\mu$. (Numerically, it is easier to 
study a lattice model than a continuum one). The Hamiltonian is 
\bea H = &-& \frac{i}{2} ~\sum_{n = 1}^{\mathcal{N}-1} ~\Big[ (\ga - \De)~
a_n b_{n+1} ~+~ (\ga+\De) ~a_{n+1} b_n \Bigl] \non \\ 
&-& \frac{i}{2}~ \sum_{n = 1}^{\mathcal{N}} ~\mu ~a_n b_n, \label{hamab}\eea
where $a,b$ are Majorana operators which are Hermitian and satisfy the 
anticommutation relations $\{a_m,a_n\}=\{b_m,b_n\}= 2\de_{mn}$ and 
$\{a_m,b_n\} = \de_{mn}$. To connect this Majorana formalism to the usual
particles, we define the particle creation and annihilation operators as 
$f_n = (1/2) (a_n - i b_n)$ and $f_n^\dg = (1/2) (a_n + i b_n)$, which 
satisfy the standard anticommutation relations $\{ f_m, f_n^\dg \} = 
\de_{mn}$. The particle number operator at site $n$ is then given by
$f_n^\dg f_n = (i a_n b_n + 1)/2$. Hence the last term in Eq.~\eqref{hamab} 
is, apart from a constant, equal to $- \mu \sum_n f_n^\dg f_n$, as a chemical 
potential term should be.

The Hamiltonian in Eq.~\eqref{hamab} has an ``effective time reversal 
symmetry", namely, it is invariant under complex conjugation of all 
complex numbers along with $a_n \to a_n$ and $b_n \to - b_n$ \cite{manisha}.
(This symmetry would be violated if terms like $i a_m a_n$ or $i b_m b_n$
were present). This symmetry implies that if we look at eigenstates with
zero energy, their wave functions involve only the $a_n$ or only the $b_n$, 
not both. Depending on the values of the parameters $\ga, ~\De$ and $\mu$, 
Eq.~\eqref{hamab} is known to have two topological phases (for $|\mu| < \ga$
and $\De \ne 0$) and one non-topological phase (for $|\mu| > \ga$). The bulk 
modes are gapped in all the phases. In the topological phases, a long system 
has one zero energy Majorana mode at each end, whose wave functions 
involve only the $a_n$ operators near the left end and only the $b_n$ 
operators near the right end if $\De > 0$, and vice versa if 
$\De < 0$~\cite{manisha}. (We can see this
particularly clearly in the special case that $\mu =0$ and $\De = \pm \ga$.
If $\De = \ga$ ($-\ga$), the Majorana modes are given by $a_1$ ($b_1$) at 
the left end and $b_{\mathcal{N}}$ ($a_{\mathcal{N}}$) at the right end).
Comparing these statements with the expressions given in the paragraph 
following Eq.~\eqref{spl} for the wave functions of Majorana modes at the ends
of a long system, we conclude that a Majorana mode made from $a_n$ ($b_n$) 
has a wave function of the form $(1,-i)^T$ ($(1,i)^T$).
 
The energies and eigenstates of Eq.~\eqref{hamab} have some interesting
properties. We can write the Hamiltonian in the general form
\beq H ~=~ i ~\sum_{mn} ~\al_m M_{mn} \al_n, \eeq
where $M_{mn}$ is a real antisymmetric matrix, and $\al_n$ denote all the 
$2 {\mathcal{N}}$ Majorana operators. The energies $E$ and corresponding 
eigenstates 
$u$ must satisfy $i M u = E u$. We then see that for every non-zero energy
$E$ and eigenstate $u$, there will be an energy $-E$ with eigenstate $u^*$.
Next, the fact that the Hamiltonian only has terms like $i a_m b_n$
implies that we can choose the eigenstates in such a way that the $a_n$ 
components are real and the $b_n$ components are imaginary. Hence, when we
go from $u$ to $u^*$, the $a_n$ components will remain the same while
the $b_n$ components change sign. This implies that the quantity $i a_n b_n$,
which is related to the particle number $f_n^\dg f_n$ at site $n$, has
opposite signs for the states $u$ and $u^*$ with energies $E$ and $-E$.

\begin{figure}[h] \ig[width=3.4in]{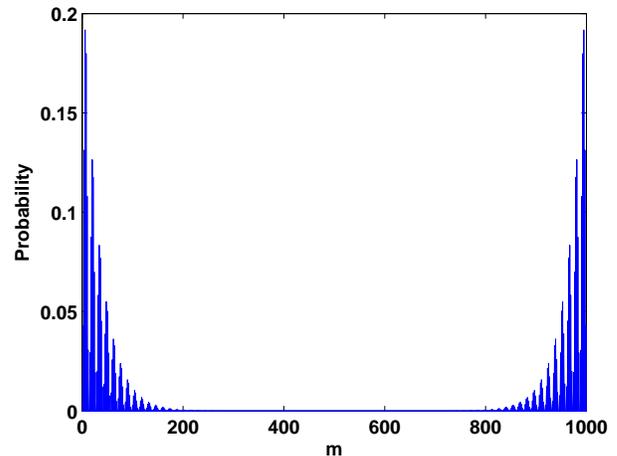}
\caption[]{(Color online) Two Majorana modes, one at each end of a 500-site 
system with $\ga=1, ~\De= 0.03$, and $\mu=0.9$.} \label{fig:maj1} \end{figure}

To numerically study the Majorana modes in this system, we first consider a 
500-site system with $\ga=1, ~\De=0.03$, and $\mu=0.9$. To distinguish 
between localized and extended states, we use the inverse participation ratio 
(IPR) \cite{mt}. (Given an eigenstate $\psi$ of the Hamiltonian, which is 
normalized so that $\sum_n |\psi_n|^2 = 1$, the IPR of the state is defined 
as $\sum_n |\psi_n|^4$). We find that for two of the eigenstates, the IPR is 
much larger than for all the other eigenstates; these correspond to localized 
states. The energy eigenvalues of these two eigenstates are zero to our
numerical accuracy. Hence we get one Majorana mode at each end of the system 
as shown in Fig.~\ref{fig:maj1}. Note that the number of components of the wave
function is twice the number of sites since each site $n$ has $a_n$ and $b_n$.

%\begin{figure}[h] \ig[width=3.4in]{NSN24.ps}

%\begin{figure}[h] \ig[width=3.4in]{NSN25.ps}
\begin{figure}[h] \ig[width=3.4in]{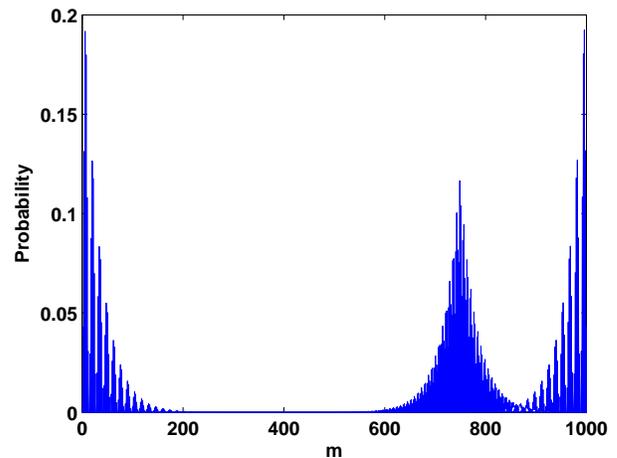}
\caption[]{(Color online) Four Majorana modes, one at each end and two in the 
middle (on two sides of the components around 700 corresponding to the site
labeled 350), for a 500-site system with $\ga=1$, $\mu=0.9$, and $\De= 0.03$ 
for sites 1 to 350 and $-0.03$ for sites 351 to 500.} \label{fig:maj2} 
\end{figure}

Next we again consider a 500-site system with $\ga=1$ and $\mu=0.9$, but now 
with $\De=0.03$ from site 1 to 350 and $\De=-0.03$ from site 351 to 500. The 
IPR is maximum for four eigenstates, indicating the presence of localized 
state. The energy eigenvalues of these four eigenstates are zero to numerical 
accuracy. The system now has four Majorana modes as shown in 
Fig.~\ref{fig:maj2}: two at the ends and two around the point where $\De$ 
changes sign. (Note that the wave function components $m=699, ~700$ correspond 
to $a_n$ and $b_n$ at the site labeled $n=350$). We note that in 
Figs.~\ref{fig:maj1} and \ref{fig:maj2}, there is no impurity potential 
anywhere inside the system.

\begin{figure}[h] \ig[width=3.4in]{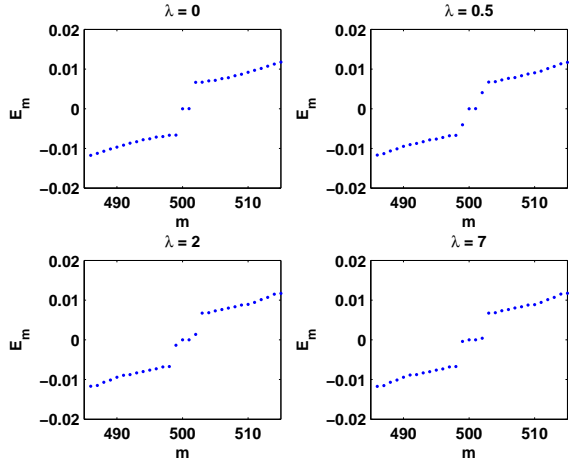}
\caption[]{(Color online) Energy levels for different values of the strength 
$\la$ of an impurity inside a SC, when $\De$ has the same value everywhere. 
The system has 500 sites, $\ga = 1, ~\mu = 0.9$, and $\De = 0.03$ at all 
sites. The impurity with strength $\la$ is at the site labeled 350. For 
small $\la$, there are only two Majorana modes, with energy close to zero, 
which lie at the ends of the SC. For large $\la$, the SC gets cut into two 
parts, and two additional Majorana modes appear near the impurity.} 
\label{fig:mm1} \end{figure}

\begin{figure}[h] \ig[width=3.4in]{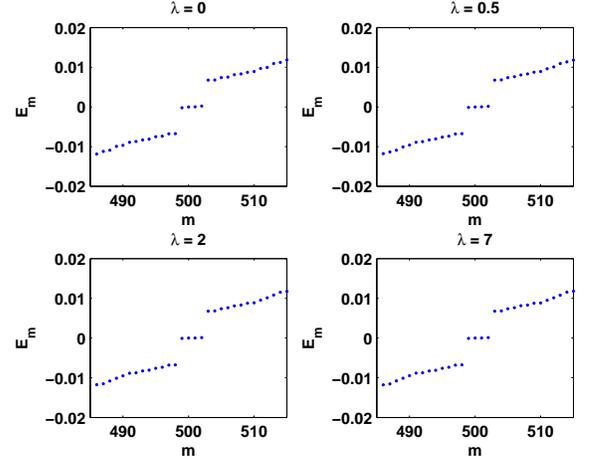}
\caption[]{(Color online) Energy levels for different values of the strength 
$\la$ of an impurity inside a SC, when $\De$ changes sign at that point.
The system has 500 sites, $\ga = 1, ~\mu =0.9$, and $\De = 0.03$ for sites 
1 to 350 and $-0.03$ for sites 351 to 500.
The impurity with strength $\la$ is at the site labeled 350. For all values of
$\la$, there are four Majorana modes with energy close to zero, two of which
lie at the ends of the SC and two are near the site where $\De$ changes sign.} 
\label{fig:mm2} \end{figure}

It is important to note that a SC in which $\De$ changes sign at one point, 
say $x=x_0$, will 
{\it necessarily} have two Majoranas near that point if the Hamiltonian 
only has terms of the form $ia_m b_n$. To see this, note that if the SC 
was cut at that point by putting an infinitely strong barrier there, 
the left part of the SC where $\De > 0$ will have a Majorana involving 
$a_n$ at its left end (i.e., at $x=-\infty$) and $b_n$ at its right end 
($x=x_0$), while the right part of the SC where $\De < 0$ will have a 
Majorana involving $b_n$ at its left end (at $x=x_0$) and $a_n$ at its 
right end ($x=\infty$). We thus see that there will be two Majoranas
near $x=x_0$ which are both of type $b$. If the barrier at $x=x_0$
is now decreased to a finite value (or even removed), the two Majoranas
will survive since the Hamiltonian has no terms of type $i b_m b_n$ which
can couple them and thereby gap them out. This is different
from a SC where $\De$ has the same sign everywhere. Then an infinitely
strong barrier at some point $x_0$ will cut the SC and produce two zero
energy Majoranas there, but these will be of opposite types, $a$ and $b$.
Lowering the barrier will now mix these Majoranas by tunneling and thus 
gap them out if the barrier is small enough. These observations are 
illustrated in Figs.~\ref{fig:mm1} and \ref{fig:mm2}.

Fig.~\ref{fig:mm1} shows some of the
eigenvalues for a 500-site system with different values of $\la$, which is
the strength of an impurity placed at the site labeled 350. The energies are 
sorted in increasing order and are labeled by $m$ which runs from 1 to 1000.
Thus $m = 500, ~501$ label the middle two energy levels; these remain 
at zero for all values of $\la$ and correspond to Majorana modes at the ends
of the SC. The Majorana modes near the site labeled 350 correspond to
$m=499, ~502$; they are at zero energy if $\la$ is large but move away
from zero and merge with the bulk states as $\la$ is decreased.

%\begin{figure}[h] \ig[width=3.4in]{NSN26.ps}

The situation is different if $\De$ changes sign at one point in the SC and 
there is also an impurity at the point. The number of Majorana modes close
to zero energy is now always four regardless of the value of $\la$; two of 
the modes lie at the ends of the system and the other two lie near the point 
where $\De$ changes sign (these modes were shown in Fig.~\ref{fig:maj2} for 
the case $\la =0$). Fig.~\ref{fig:mm2} shows the energies of these four modes 
for a 500-site system with $\De$ changing sign and an impurity of strength 
$\la$ at the site labeled 350. We see that the four energies lie close to 
zero for all $\la$, unlike Fig.~\ref{fig:mm1} where that happens only if 
$\la$ is large.

%\begin{figure}[h] \ig[width=3.4in]{NSN27.ps}

%\begin{figure}[h] \ig[width=3.in]{NSN1a.ps}
\begin{figure}[h] \ig[width=3.in]{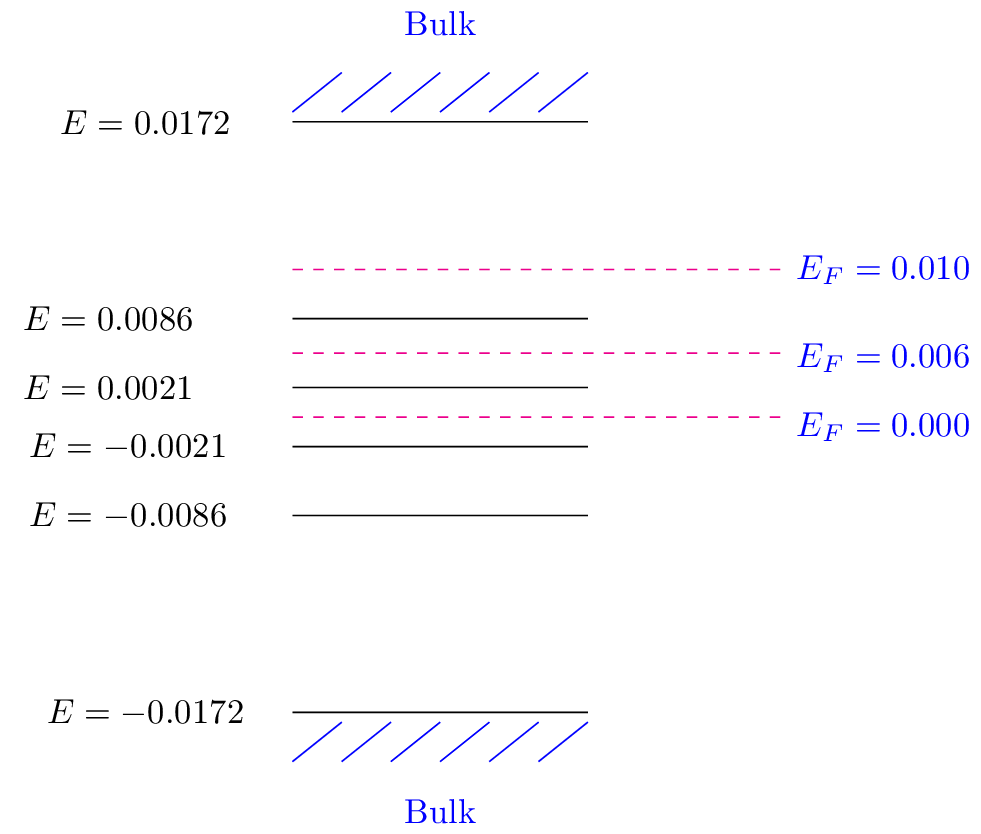}
\caption[]{(Color online) Energy spectrum for a 60-site system with $\ga=1$, 
$\mu=0.9$, and $\De=0.03$ from sites 1 to 45 and $-0.03$ from sites 45 to 60. 
The three values of $E_F$ shown correspond to the three plots in 
Fig.~\ref{fig:pd}.} \label{fig:energy} \end{figure}

We will now look at the effect of the Majorana modes on the local particle 
density. Let $a_{E,n}$ and $b_{E,n}$ denote the $a$ and $b$ components of
an eigenstate with energy $E$. For a Fermi energy $E_F$, all energy levels 
up to $E_F$ will be filled. We then define the total particle density at 
each site $n$ as 
\beq \rho_n ~=~ \frac{1}{2} \sum_{E < E_F}^{} (ia_{E,n} b_{E,n} ~+~ 1), 
\label{rho} \eeq
where $E_F$ is the Fermi energy of the system and $n$ goes from 1 to 
$\mathcal{N}$. We consider a system of size $\mathcal{N}= 60$, $\ga=1, ~\mu
=0.9$, with $\De$ equal to $0.03$ from sites 1 to 45 and $-0.03$ from sites 
45 to 60. This system size is not very large, so the four Majorana modes
will hybridize with each other and their energies will split from zero.
We find numerically that the energies of the four modes lie
at $\pm 0.0086$ and $\pm 0.0021$ as shown in Fig.~\ref{fig:energy}. 
The bulk gap is found to be $0.0344$, so these four energies lie well within 
the bulk gap. We now study the particle density given in Eq. (\ref{rho})
for three values of Fermi energy $E_F= 0.010, 0.006, 0.000$ to see the effect 
of the Majorana modes on the particle density. For $E_F=0.010$, the PD does 
not get any contribution from Majorana modes as the contributions to 
$ia_n b_n$ from pairs of states with energies $\pm E$ cancel out. 
For $E_F=0.006$, the PD gets contribution only from one unpaired Majorana 
mode (with $E=-0.0086$). For $E_F=0.000$, PD gets contribution from both 
the unpaired Majorana modes ($E=-0.0086$ and $-0.0021$). The PD for 
these three values of $E_F$ is shown in Fig.~\ref{fig:pd}. 
A comparison of the three plots will show the contribution of the Majorana
modes to the PD; for instance, the difference of the PD in plots (a) and (b)
comes from the Majorana mode at $E=-0.0086$ while the difference of the PD 
of the plots (b) and (c) comes from the Majorana mode at $E=-0.0021$.
 
\begin{widetext}
\begin{center} \begin{figure}[h]
%\begin{center} \begin{tabular}{ccc}
%\epsfig{figure=NSN17.ps,width=2.2in,height=1.87in,clip=} &
%\epsfig{figure=NSN18.ps,width=2.2in,height=1.87in,clip=} &
%\epsfig{figure=NSN19.ps,width=2.2in,height=1.87in,clip=} \\
\subfigure[]{\ig[width=2.2in]{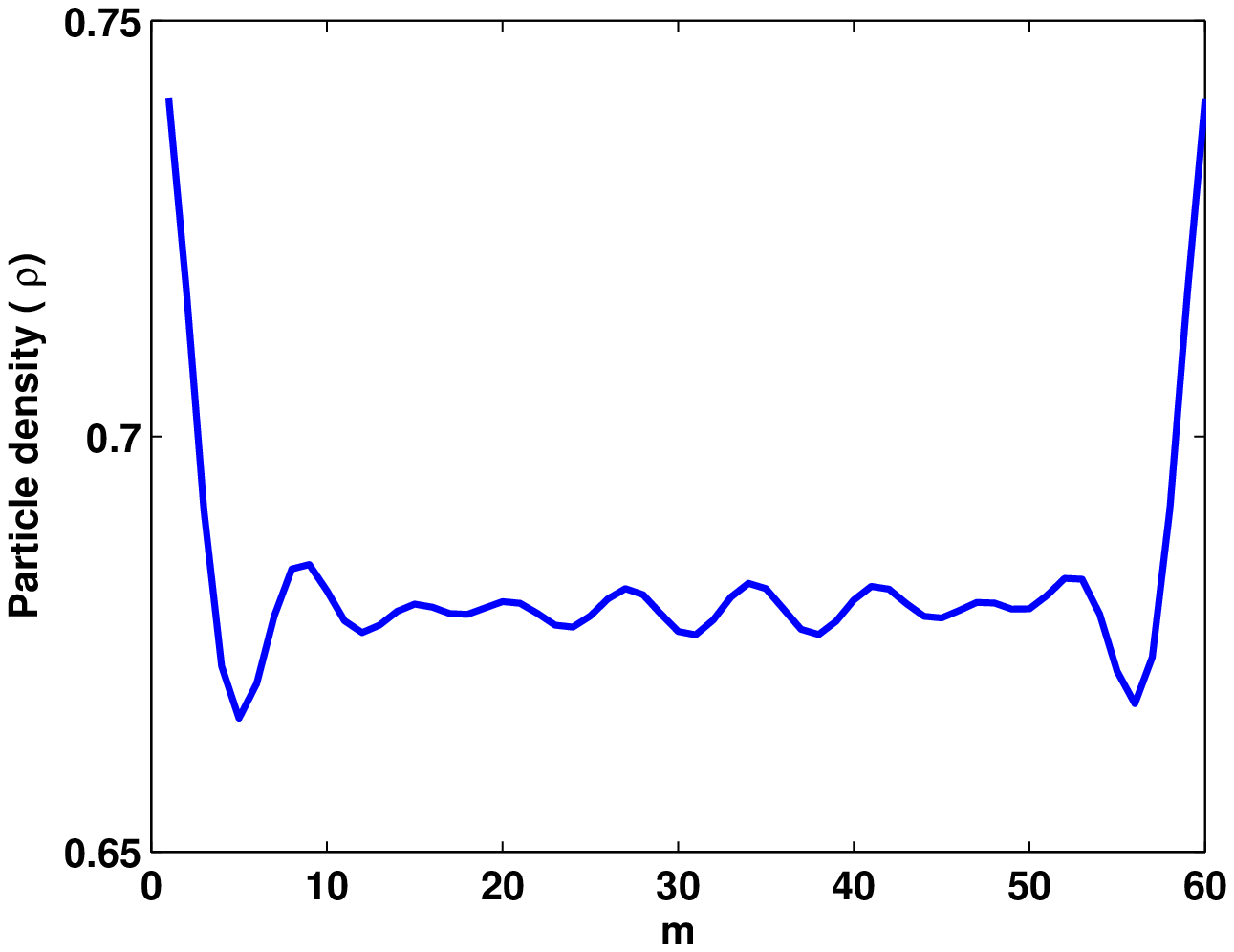}}
\subfigure[]{\ig[width=2.2in]{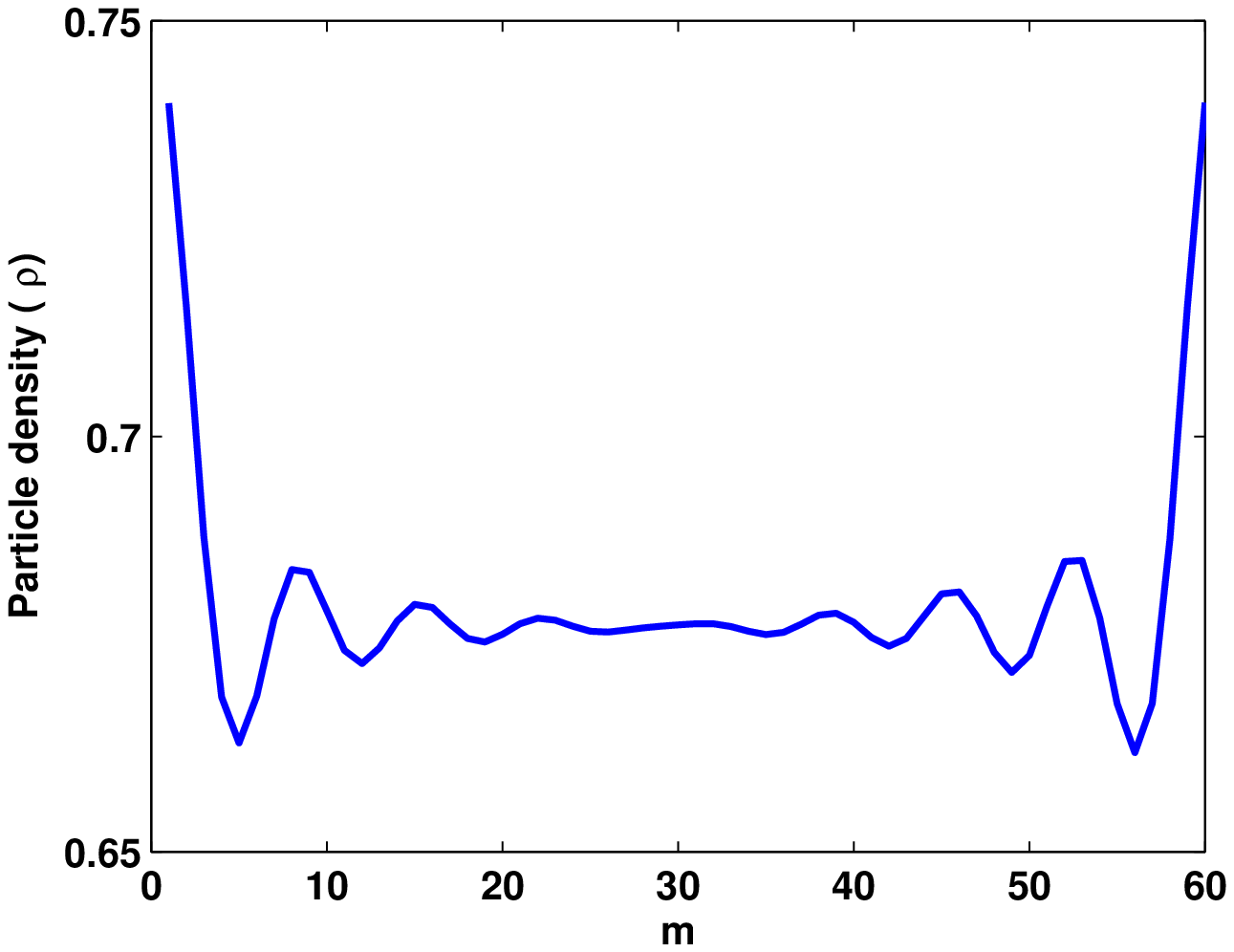}}
\subfigure[]{\ig[width=2.2in]{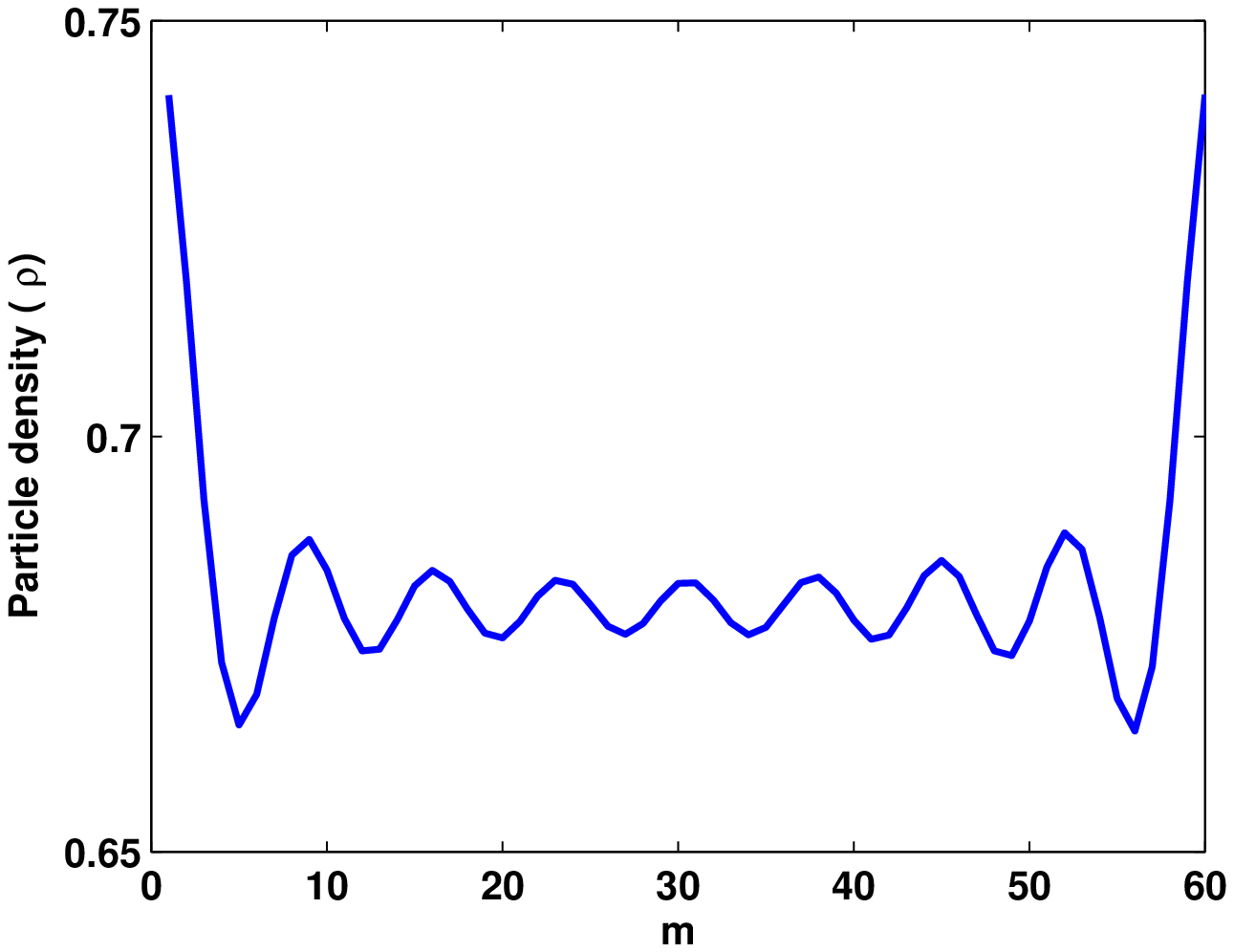}} \\
%\end{tabular} \end{center}
\caption{(Color online) Plots of the particle density when $\De$ changes sign 
at one point in the SC. We take the
the SC region to have 60 sites, with $\ga=1$, $ \mu=0.9$, $\De=0.03$ from 
sites 1 to 45, and $\De=-0.03$ from sites 45 to 60. As the system length
is finite, the four Majorana modes (two near the ends and two near the $45$-th
site) split from zero energy. The energies of the four modes are $\pm 0.0086$
and $\pm 0.0021$. Figure (a) is for $E_F = 0.010$ which lies above all 
the Majorana energies; hence the PD does not get a contribution from any of 
the Majorana modes since they are all occupied but their contributions cancel 
out in pairs. Figure (b) is for $E_F = 0.006$; here the PD gets a 
contribution only from the Majorana mode at $E= - 0.0086$ since the mode at 
$E=0.0086$ is unoccupied and therefore does not contribute. Figure (c) is for
$E_F = 0.000$; now the PD gets contributions from the Majorana modes at 
both $E=-0.0086$ and $-0.0021$ since the modes at $E=0.0086$ and $0.0021$ are
unoccupied and do not contribute.} \label{fig:pd} \end{figure}
\end{center}
\end{widetext}

\section{Interactions and renormalization group analysis}
\label{sec:rg}

We will now study the effect of interactions between the electrons on the
Majorana modes. More precisely, we will use the RG method
to study how the various parameters of the model vary with the length scale,
and we will then use that to see what happens to the Majorana modes.
To begin, we will study only the SC region and not the NM leads. At the
end of this section, we will consider the RG equation for a tunnel
barrier lying at the junction of a SC and a NM.

Interactions have a particularly strong effect on many-electron systems 
in one dimension. Short range interactions change the system from a Fermi 
liquid to a Tomonaga-Luttinger liquid (TLL). The most effective way to study 
a TLL is to use bosonization~\cite{gogolin,delft,rao,giamarchi}. Let us 
briefly explain the bosonization formalism. We first expand the second 
quantized electron fields around the Fermi wave numbers $\pm k_F$ as 
\beq \Psi ~=~ e^{ik_F x} ~\Psi_R ~+~ e^{-ik_F x} ~\Psi_L, \eeq
where $\Psi_R$ and $\Psi_L$ denote right and left moving linearly dispersing 
fields. In bosonization these are related to two conjugate bosonic fields 
$\phi$ and $\ta$ as
\beq \Psi_{R/L} ~\sim~ \frac{1}{\sqrt {2\pi a_0}} ~\exp [i \sqrt{\pi} ~(\mp 
\phi + \ta)], \label{bos} \eeq
where $a_0$ is a microscopic length scale such as the lattice spacing
or the distance between nearest neighbor particles. (For simplicity we are
ignoring Klein factors in Eq.~\eqref{bos}).
The fields $\phi$ and $\ta$ describe particle-hole excitations, and their
space-time derivatives give the charge current $J_c$ and the deviation of the 
charge density from a uniform background density, $\de \rho = \rho_c - \rho_0$.
Short range density-density interactions of the form $\int \int dx dy 
\rho(x) V_{int} (x-y) \rho (y)$ are therefore quadratic in terms of $\phi$ and
$\ta$; this is the key advantage of the bosonization formalism. The Dirac 
Hamiltonian $H = i v_F \int dx [ - \Psi_R^\dg \pa_x \Psi_R + \Psi_L^\dg \pa_x 
\Psi_L]$ (here $v_F$ is the Fermi velocity of the non-interacting system of 
electrons) along with density-density interactions takes the bosonic 
form $H= (v/2) \int dx [ K (\pa_x \ta)^2 + (\pa_x \phi)^2 /K]$, where
$v$ is the velocity of the particle-hole excitations in the interacting theory,
and $K$ is a dimensionless parameter called the Luttinger parameter; these
are related to $v_F$ and the strength of the interactions. [For repulsive 
(attractive) interactions between the electrons, $K<1$ ($K>1$)].
The superconducting term $\Psi_R^\dag \Psi_L^\dg$ plus its Hermitian conjugate
is proportional to $\sin (2 \sqrt{\pi} \ta)$. Finally, if we have a lattice 
model which is at half filling, there will be umklapp scattering terms like
$\Psi_R^\dg \Psi_R^\dg \Psi_L \Psi_L e^{-i4k_F x}$ plus its Hermitian 
conjugate which add up to $\cos (4 \sqrt{\pi} \phi - 4 k_F x)$. Putting all 
this together, we get a bosonized Hamiltonian of the form 
\cite{giamarchi,ganga}
\bea H &=& \int \frac{dx}{2} [ ~v K (\pa_x \ta)^2 ~+~ \frac{v}{K} 
(\pa_x \phi)^2 ~+~ \frac{4\de}{\pi a_0^2}~\sin(2 \sqrt{\pi} \ta) \non \\
&& ~~~~~~~~~- ~\frac{U}{\pi^2 a_0^2}~\cos(4 \sqrt{\pi} \phi - 4 k_F x) ], \eea
where $\de$ is related to the SC pairing $\De$ as $\de = \De a_0$ at
the microscopic length scale (we will see below that all these quantities
will change with the length scale). We will ignore the umklapp scattering 
term below by setting $U=0$.

Let us consider the case where there is an isolated impurity in the system;
for simplicity, we will assume this to be point-like so that the impurity
potential is $V(x)= \la \de(x)$, $\la$ being the strength of the impurity. 
The Hamiltonian which describes the effect of this is given by
\bea H_{imp} ~=~ \int dx ~V(x)\rho(x) ~=~ \la \rho(0), \non \\ \eea
where the density $\rho(x) = -(1/\pi) \pa_x \phi(x)$. The interaction 
renormalizes the system parameters. Hence the RG equations for the length 
scale dependence of the parameters $K$, $\de$ and $\la$ are given by
\bea \frac{dK}{dl}&=& \frac{\de^2}{2}, \non \\
\frac{d \de}{dl}&=&(2- \frac{1}{K}) ~\de, \non \\
\frac{d\la}{dl}&=&(1- K) ~\la. \label{rg0} \eea
%(The velocity $v$ also flows under RG but we will ignore this here).
It is convenient to define a renormalized length scale $a= a_0 e^l$. In the 
figures below, we will plot the pairing $\De$ which is related to $\de$
as $\De = \de/a$; this satisfies the equation
$d\De /dl = (1-1/K) \De$. It is $\De$, rather than $\de$, which is physically 
observable; for instance $\De$ is the superconducting gap for a system with 
length scale $a$. Note that for a non-interacting system, i.e., $K=1$, both 
$\de$ and $a$ flow, but $\De$ does not flow. We note that the first two 
equations in Eqs.~\eqref{rg0} were studied in Ref. \onlinecite{ganga}. We 
have generalized their analysis by introducing an impurity with strength 
$\la$ which flows according to the last equation in Eqs.~\eqref{rg0}. It 
is important for us to consider this equation since we are mainly interested 
in the conductances of the system and these are strongly affected by the 
presence of impurities or barriers (see Eq.~\eqref{rg3}). As we will see 
below, an impurity inside a SC region can have interesting consequences for 
Majorana modes and the conductances.

\begin{figure}[h] \ig[width=3.4in]{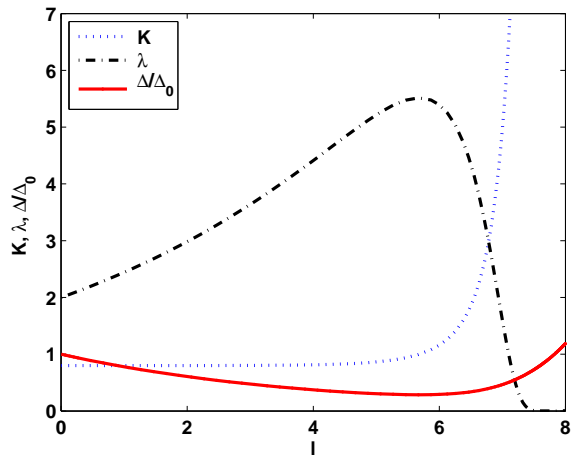}
\caption[]{(Color online) Plots of $K$, $\De/ \De_0$ and $\la$ versus $l$. 
The initial values of the parameters are $\De_0=0.01$, $K_0=0.8$ and $\la_0=
2$.} \label{fig:rg1} \end{figure}

\begin{figure}[h] \ig[width=3.4in]{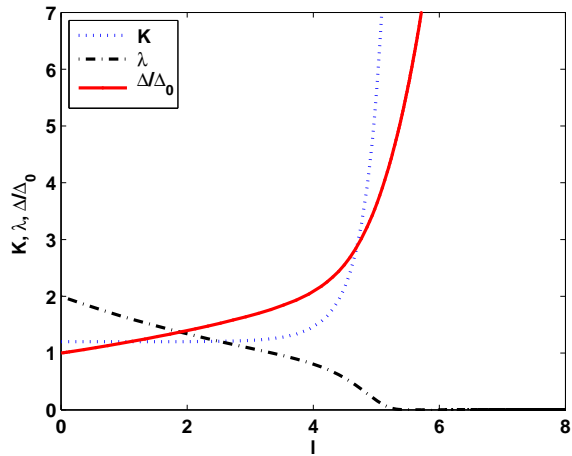}
\caption[]{(Color online) Plots of $K$, $\De/ \De_0$ and $\la$ versus $l$. 
The initial values of the parameters are $\De_0=0.01$, $K_0=1.2$ and $\la_0
=2$.} \label{fig:rg2} \end{figure}

We have used the RG equations and some initial values of the parameters 
$K$, $\De/ \De_0$ and $\la$ to find how these parameters vary with the length 
scale. These are shown in Figs.~\ref{fig:rg1} and \ref{fig:rg2} for initial
values $K_0 = 0.8$ (repulsive interactions) and $K_0 = 1.2$ (attractive 
interactions) respectively, along with $\De_0=0.01$ and $\la_0= 2$. For $K<1$,
$\la$ increases and $\De$ decreases with increasing length scale,and these 
trends are reversed when $K>1$. In Fig.~\ref{fig:rg1}, where $K_0 = 0.8$,
we see that $\la$ keeps increasing up to $l=6$ (beyond this $K$ becomes larger
than 1). These results have the following implications for Majorana modes
at the ends of the SC.

%\begin{figure}[h] \ig[width=3.4in]{NSN20.ps}

%\begin{figure}[h] \ig[width=3.4in]{NSN22.ps}

If there is no impurity inside the SC (i.e., $\la = 0$), then we will be in 
the situation studied in Ref. \onlinecite{ganga}. If the SC region has a
length $L$, the Majorana modes survive if the value reached by $\De$ at that
length scale satisfies $L \De/v \gg 1$. If $L \De/v \lesssim 1$, the Majorana 
modes will hybridize strongly and move away from zero energy; if their
energies approach the ends of the SC gap at $\pm \De$, they will become 
unobservable.

If $\la \ne 0$, it will grow with the length scale if $K < 1$. For a large 
impurity strength $\la$ at one point, the SC essentially gets cut into two 
parts and one can get two Majorana modes on the two sides of that 
point~\cite{stanescu4}. As shown in Fig.~\ref{fig:mm1} for a 500-site system, 
the energies of these Majorana modes approach zero as $\la$ becomes large.
(As discussed below, the RG equations cannot be trusted up to very large
values of $\la$. The largest value $\la =7$ shown in Figs.~\ref{fig:mm1}
and \ref{fig:mm2} is therefore only for the purposes of illustration and
may not be physically realistic).
The situation is different if $\De$ changes sign at one point in the SC and 
there is also an impurity at the point. The RG equations will be the same as 
discussed above for the case of uniform $\De$. However the number of Majorana 
modes is now always four regardless of the value of $\la$. Two of the modes 
lie at the ends of the system and the other two lie near the point where $\De$ 
changes sign; the energies of these four modes are shown in Fig.~\ref{fig:mm2}.
We see that the four energies lie close to zero for all $\la$. 

We have studied above what happens if there is an impurity of strength $\la$
which lies inside the SC. It is also interesting to study the RG flow of an 
impurity which lies at the ends of the SC region, namely, the barriers which 
lie at the junctions between the SC and a NM lead. It is known that an 
impurity of strength $\la$ lying at the junction between two different TLLs 
with Luttinger parameters $K_1$ and $K_2$ satisfies the RG equation
\beq \frac{d\la}{dl} ~=~ (1 ~-~ \frac{2K_1 K_2}{K_1 ~+~ K_2}) ~\la. \eeq
Since a NM is equivalent to a TLL with $K=1$, the RG equation for the
strength of an impurity at the junction between a NM and a SC which has a 
value of $K$ is given by
\beq \frac{d\la}{dl} ~=~ \frac{1 ~-~ K}{1 ~+~ K} ~\la. \label{rg3} \eeq
If $K < 1$, we see that a barrier strength $\la$ will increase with the 
length scale, though not as fast as the $\la$ of an impurity lying inside 
the SC as shown by the last equation in Eq.~\eqref{rg0}.

We can apply Eq.~\eqref{rg3} to understand the conductance across a NSN 
system if we take $\la$ to be the strength of the tunnel barriers between 
the SC and the NM leads. If both $\la$ and $e^{L/\xi}$ are large and 
$k_F L$ is not an integer multiple of $\pi$, Eqs.~\eqref{nu} give 
expressions for the conductances across a NSN system at zero bias. 
For $E=0$, $L/\xi = L \De/v$; hence the parameter $\nu$ 
in Eqs.~\eqref{nu} depends on $\la$ and $\De$ both of which flow under
RG. If $K < 1$, $\la$ increases and $\De$ decreases with increasing length 
scales; both these imply that $\nu$ will increase and hence $G_C$ will 
decrease as the length $L$ of the SC is increased.

We would like to emphasize here that we have only discussed RG equations 
up to the lowest possible order in $\de$ and $\la$. Hence the RG flows cannot 
be trusted when these parameters reach values of the order of the energy 
cut-off, namely, the Fermi energy. Hence we cannot definitely conclude 
that the RG flow will cut the wire into two parts. However, we can conclude
that an impurity inside the superconducting part of the system will grow 
and may thereby give rise to additional sub-gap modes near that point in 
the case where the $p$-wave pairing $\De$ has the same sign everywhere. 
(If $\De$ changes sign at one point, two additional Majorana modes will
appear there regardless of whether or not there is a barrier there, as 
we have argued on symmetry grounds).

\section{Experimental realizations}
\label{sec:expt}

We will now discuss how the different systems that we have discussed above
can be experimentally realized. In particular, we will see how it may be 
possible to have a SC in which the pairing $\De$ changes sign at one point.

We consider the model studied in Ref. \onlinecite{ganga}. We take
a wire with a Rashba spin-orbit coupling of the form $\pm \al_R p_r \si^x$, 
where $p_r$ is the momentum along the wire and $\si^x$ is a Pauli spin matrix. 
This form can be justified as follows. Let us take the coordinate in the wire 
to increase along an arbitrary direction $\hat r$ lying in the $x-y$ plane 
(instead of the $\hat x$ direction as assumed in earlier sections). If the 
Rashba term is 
$\al_R {\hat n} \cdot {\vec \si} \times {\vec p}$, and $\hat n$ points in the 
$\hat z$ direction, then the Rashba term will be $\al_R p_r \si^x$ if 
${\hat r}={\hat y}$ and $- \al_R p_r \si^x$ if ${\hat r}=-{\hat y}$.

Next, the wire is placed in a magnetic field in the $\hat z$ direction 
which is perpendicular 
to the Rashba term (with a Zeeman coupling $\De_Z$) and in proximity to 
a bulk $s$-wave SC with pairing $\De_S$. The complete Hamiltonian, given in 
Ref. \onlinecite{ganga}, is
\bea H &=& \int dr ~\Psi^\dg_\al \left[ (\frac{p_r^2}{2m} - \mu)\de_{\al \be} 
\pm \al_R p_r \si^x_{\al \be} - \De_Z \si^z_{\al \be} \right] \Psi_\be \non \\
&& + \frac{i}{2} \int dr ~[\De_S \Psi^\dg_\al \si^y_{\al \be} \Psi^\dg_\be +
H.c.], \label{rashba} \eea
where $\Psi_\al$ is the annihilation operator for an electron with spin $\al$,
and the $\pm$ sign of the Rashba term depends on whether ${\hat r} = \pm 
{\hat y}$. Ref. ~\onlinecite{ganga} then shows that for a 
certain range of the parameters, this system is equivalent to a spinless 
$p$-wave SC of the form that we have studied in this paper, with the $p$-wave 
pairing being given by $-i (\De/k_F) (c^\dg \pa_x d + d^\dg \pa_x c)$ (see 
Eq.~\eqref{Ham}), where 
\beq \De ~=~ \pm ~\frac{\hbar \al_R k_F \De_S}{\De_Z}. \label{eff} \eeq

\begin{figure}[h] \ig[width=2.5in]{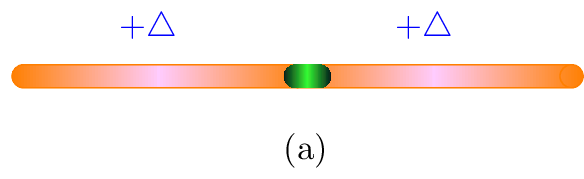} \\ \ig[width=1.5in]{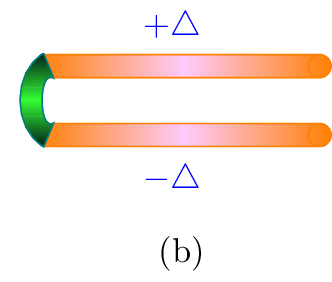}
\caption[]{(Color online) Superconducting systems with (a) the same value of 
$\De$ throughout, and (b) the sign of $\De$ changing near one point shown in 
green (dark shade). There is also an impurity present at that point.} 
\label{exp} \end{figure}

Now consider a straight wire in which the coordinate ${\hat r} ={\hat y}$
along the entire wire; see Fig.~\ref{exp} (a). Then the Rashba term and 
hence $\De$ will have the same sign everywhere. On the other hand, suppose 
that the wire is bent by an angle $\pi$ so that the two parts of the wire 
run in opposite directions as shown in Fig.~\ref{exp} (b). Now ${\hat r} = 
-{\hat y}$ in the lower part of the wire and ${\hat r} = {\hat y}$ in the 
upper part. Then Eq.~\eqref{eff} shows that $\De$ will have opposite signs in 
the two parts. It is also clear that the bend in the wire is likely to cause 
some scattering of the electrons, and it is natural to model such a scattering
by assuming that an impurity potential is present there.

Note that if the wire is bent by any angle different from zero or $\pi$,
the situation will be more complicated because the Rashba term
$\al_R {\hat n} \cdot {\vec \si} \times {\vec p}$ will no longer be 
proportional to the same $\vec \si$ matrix in the two parts of the wire.
Hence the effective $p$-wave superconductors in the two parts will not
be related simply by a phase change in $\De$.

%\begin{figure}[h] \ig[width=2.5in]{NSN1c.ps} \\ \ig[width=1.5in]{NSN1d.ps}

The impurity inside the SC that we have studied in Sec. \ref{sec:rg} 
corresponds to an arbitrary point in Fig.~\ref{exp} (a) and to the region 
where $\De$ changes sign in Fig.~\ref{exp} (b); both these points are shown 
in green (dark shade). The strength of the impurity $\la$ can be controlled 
by placing a gate near those points and varying the gate voltage.

Finally, the ends of the SC can be connected to NM leads through tunnel
barriers. As discussed in Sec. \ref{sec:model}, these barriers can be 
characterized by their strength $\la$.

In Secs. II-IV, we discussed a conductance $G_C$ in which pairs of 
electrons can appear in (or disappear from) the $p$-wave SC. 
At a microscopic level we can understand these processes as occurring due 
to a Cooper pair going from the $s$-wave SC to the $p$-wave 
SC (or vice versa). Finally we have to assume that the two
NM leads and the $s$-wave SC form a closed electrical 
circuit so that we can measure the conductances $G_N$ and $G_C$.

We would like to note here that some other realizations of systems in which 
the pairing $\De$ changes sign at one point have been discussed 
before~\cite{alicea1,kesel,ojanen,lucig,klino3}. However the effect of an 
impurity at that point and the conductances of the system were not studied in 
these papers. Although the geometry of the system discussed in Ref. 
\onlinecite{kesel} looks similar to ours (Fig.~\ref{exp} (b)), the details 
are quite different. The set-up proposed in Ref. \onlinecite{kesel} requires 
proximity to two different $s$-wave superconductors and also requires a 
superconducting loop threaded by a magnetic flux equal to $hc/4e$.

\section{Conclusions}
\label{sec:concl}

In this paper, we have studied the Majorana modes and conductances of a 
one-dimensional system consisting of a $p$-wave SC of length $L$ connected 
by tunnel barriers to two NM leads. We have considered two cases: ~(i) the 
$p$-wave $\De$ pairing has the same sign everywhere in the SC (with, possibly,
an impurity potential with strength $\la$ present at one point inside the SC),
and (ii) $\De$ changes sign at one point in the SC (and an impurity present 
there). We have used a continuum model to derive the boundary conditions at 
the junctions between the SC and the NM. Using these conditions, we have 
numerically studied two conductances, $G_N$ (from one lead to the other) and 
the Cooper pair conductance $G_C$ (from one lead to the SC), when the energy 
$E$ of an electron incident from one of the leads lies within the SC gap. 
% ($E$ can also be interpreted as the bias between that lead and the SC). 

We have studied three ranges of values of $L$ with respect to the length 
$\eta$ (the length scale associated with the SC gap).
We find a rich pattern of variations of the 
conductances as functions of $L$ and $E$. We have provided analytical 
explanations for these behaviors by studying some special limits, such as 
the Majorana modes at the ends of a SC box with no leads, and the 
conductances of the NSN system in the limit when the tunnel barriers $\la$ 
between the SC and the leads are very large. In this limit, we find that there
are quantization conditions for the length $L$ at which $G_N$ and $G_C$ 
have peaks. We find that the presence of Majorana modes at the ends
of the SC has a significant effect on the conductances; the latter have
peaks exactly at the energies of the Majorana modes. We do not find any 
noticeable difference between the behaviors of the conductances for the 
cases of uniform $\De$ versus $\De$ changing sign at one point in the SC,
although the latter system has two additional Majorana modes at that point. 
This implies that the presence of Majorana modes inside the SC (i.e.,
far away from the leads) has no major effect on the conductances.

[We would like to mention here that the hybridization of the Majorana 
modes at the ends of the wire can also occur due to tunneling processes 
involving virtual quasiparticle states in the bulk $s$-wave SC which is in 
proximity to the wire~\cite{zyuzin}. This can give rise to an energy splitting 
even if the Majorana modes cannot directly hybridize through the wire. A 
discussion of this effect is beyond the scope of our model].

For the case that $\De$ changes sign at one point in the SC, we have used a 
lattice model to study the Majorana modes which occur near that point. We 
find that these modes have a noticeable effect on the local particle density 
as a function of the Fermi energy. Further, these modes are very robust in 
that they stay at zero energy even if we vary the potential near that point.
This is because a symmetry of the system prevents these modes from hybridizing
with each other.

Next, we have studied the effect of interactions between the electrons in a 
SC. Using bosonization, we have studied the RG flows of the different 
parameters of the theory such as the pairing $\De$ and the strength $\la$ 
of an impurity potential which may be present either inside the SC or at the 
junctions of the SC and the leads as tunnel barriers. For repulsive
interactions, the Luttinger parameter $K < 1$; we then find that $\De$
decreases while $\la$ increases as the length scale increases. We studied
the effect of this on the Majorana modes and on the conductances. In 
particular, we find that if an impurity is present inside the SC, it
can grow and eventually cut the SC into two parts if $L$ is large enough;
then two Majorana modes can appear near the impurity. This is in contrast
to the case where $\De$ changes sign at one point and there is also an
impurity present there. We then find that there are always two Majorana modes 
near that point regardless of how small or large the impurity strength is.

Finally, we have discussed some experimental implementations of our model.
We have shown that the cases of both uniform $\De$ and $\De$ changing at
one point can be realized. The second case is interesting because two
Majorana modes are expected to appear near that point. We have shown 
that these additional modes have no noticeable effect on the conductances of
the system; this may be because we have considered a configuration 
in which the NM leads lie far away from these modes. It should be possible 
to study these modes by attaching a lead at that point and measuring the 
conductance in that lead \cite{zhou,weit}. Another way to detect these 
modes would be through STM studies of the local particle density as a 
function of energy. 

% We have presented the analytical and 
% numerical solutions for the measurement of $G_N$ and $G_C$ across 
% the junction. We have analytically solved the boundary conditions at 
% the junctions and plotted $G_N$ and $G_C$ for different values of 
% $E/\De$, and $L$. We consider three value of $L$, namely when $L\ll \eta$, 
% $L\sim \eta$, and $L\gg \eta$. We then numerically showed that for small 
% (large) $L$, only $G_N$ ($G_C$) is non-zero. For special parameter values, 
% we have solved 8 boundary conditions and 
% calculated the exact form of $r_a$, $t_n$, and quantization condition of 
% $k_F L$ with an offset. We have calculated the exact form of variation of 
% $E/\De$ with $L$ for a superconducting box with infinite wall. These 
% analytic results match exactly with the numerical results.

% Next we have examined the effect of changing sign of superconducting gap on 
% the conductivity. For this case, we have solved 12 boundary conditions,
% and have shown that 
% $G_N$ and $G_C$ follow the similar behavior as the system without $\De$ 
% changing follows. At the point of transition of $\De$, we have numerically 
% confirmed the presence of Majorana mode in a $p$-wave superconductor on a 
% lattice by two ways: (i) For large $L$, calculated the Majorana wave 
% function explicitly, (ii) For small $L$, calculated the PD. 

\section*{Acknowledgments}
We thank S. Das, J. N. Eckstein, T. Giamarchi, S. Hegde, T. L. Hughes, D. Loss,
S. Rao, K. Sengupta, A. Soori and S. Vishveshwara for stimulating discussions.
For financial support, M.T. thanks CSIR, India and D.S. thanks DST, India for 
Project No. SR/S2/JCB-44/2010.

\end{document}